W
\documentclass[twocolumn,aps,pre,epsfig,amsmath,eqsecnum]{revtex4}

\usepackage{dcolumn}
\usepackage[dvips]{graphicx}
\newcommand{\Rv}{{\vec R}}
\newcommand{\rv}{{\mathbf r}}
\newcommand{\ev}{{\hat{e}}}
\newcommand{\hv}{{\vec h}}
\newcommand{\uv}{{\mathbf u}}
\newcommand{\qv}{{\mathbf q}}

\newcommand{\nv}{{\mathbf n}}
\newcommand{\Tr}{{\rm Tr}}
\newcommand{\px}{{\partial_x}}

\newcommand{\ppi}{{\partial_i}}
\newcommand{\ppj}{{\partial_j}}

\newcommand{\calH}{{\mathcal H}}
\newcommand{\calF}{{\mathcal F}}
\def\mm#1{\underline{\underline{{#1}}}}
\def\um{{\mm{u}}}
\def\Lm{{\mm{\Lambda}}}

\def\Im{{\mm{I}}}

\def\O{{\mathcal{O}}}

\begin{document}
\title{Fluctuating Nematic Elastomer Membranes: a New Universality Class}
\author{Xiangjun Xing$^{(1)}$, Ranjan Mukhopadhyay$^{(2)}$\footnote{Current 
Address: NEC Laboratories America, Inc., 4 Independence Way, Princeton, 
NJ 08540.},
T. C. Lubensky$^{(2)}$, and Leo Radzihovsky$^{(1)}$}
\address{$(1)$ Department of Physics, University of Colorado,
Boulder, CO 80309}
\address{$(2)$ Department of Physics, University of Pennsylvania,
Philadelphia, Pennsylvania 19174}
\date{\today}
\begin{abstract}
  We study the flat phase of nematic elastomer membranes with
  rotational symmetry spontaneously broken by in-plane nematic order.
  Such state is characterized by a vanishing elastic modulus for
  simple shear and soft transverse phonons.  At harmonic level,
  in-plane orientational (nematic) order is stable to thermal
  fluctuations, that lead to short-range in-plane translational
  (phonon) correlations.  To treat thermal fluctuations and relevant
  elastic nonlinearities, we introduce two generalizations of
  two-dimensional membranes in a three dimensional space to arbitrary
  $D$-dimensional membranes embedded in a $d$-dimensional space, and
  analyze their anomalous elasticities in an expansion about $D=4$. We
  find a new stable fixed point, that controls long-scale properties
  of nematic elastomer membranes. It is characterized by singular
  in-plane elastic moduli that vanish as a power-law
  $\eta_{\lambda}=4-D$ of a relevant inverse length scale (e.g.,
  wavevector) and a finite bending rigidity.  Our predictions are
  asymptotically exact near $4$ dimensions.
\end{abstract}
\maketitle
\section{Introduction}
\label{intro}

The ubiquity and importance of membrane realizations in nature, as
cellular walls, and in laboratories, as self-assembled bilayers of
lipid amphiphils, has fueled considerable scientific
activities\cite{JWS}. These nearly tensionless\cite{tensionless} lipid
sheets are highly flexible, with elastic moduli often comparable to
thermal energies. Consequently, on long length scales, their
conformational properties are strongly affected by thermal
fluctuations. This, together with early successes in understanding a
variety of puzzling phenomena (such as for example red blood cells'
flicker\cite{BouchaudLennon}, biconcave shape of
eurythrocites\cite{Helfrich}, and period of lyotropics) in terms of
continuum models of fluctuating elastic sheets, has attracted the
attention of the physics community.  Consequently, significant
progress has been made in understanding the statistical mechanics of
fluctuating membranes\cite{JWS}

It is by now well-appreciated that the nature of a membrane's in-plane
order, with three (heretofore studied) universality classes, the
isotropic, hexatic and solid (i.e., tethered or polymerized),
crucially affects its conformational properties. The most striking
effect of in-plane orders is the stabilization in solid membranes of a
``flat" phase\cite{NP}, with long-range orientational order in the
local membrane normals\cite{AGL}, that is favored at low-temperature
over the entropically preferred high-temperature crumpled state.
Therefore, in marked contrast to liquid membranes and one-dimensional
polymer analogs, which are always crumpled (beyond a persistence
length)\cite{persistent_length}, tethered membranes, despite of being
two-dimensional\cite{MerminWagner} are predicted\cite{NP} to undergo a
thermodynamically sharp crumpled-to-flat phase
transition\cite{KKN,PKN}.  The ordering is made possible by a subtle
interplay of thermal fluctuations with nonlinear membrane elasticity,
which, at long scales infinitely enhances a membrane's bending
rigidity, thereby stabilizing the orientational order against these
very fluctuations. This novel ``order from disorder'' phenomenon, and
the universal ``anomalous elasticity'', namely, length-scale dependent
elastic moduli, non-Hookean stress-strain relation, and a universal
negative Poisson ratio\cite{NP,AGL,GD,LR}, are now known to be quite
commonly exhibited by many other ``soft'' systems subjected to
fluctuations.

In this paper we introduce and explore a new universality class of
{\em solid} membranes, that {\em spontaneously} develop an {\em
  in-plane} orientational nematic order.  Our motivation is two-fold.
Firstly, our interest is driven by experimental progress in the
synthesis of nematic liquid crystal elastomers\cite{review_exp},
statistically isotropic and homogeneous gels of crosslinked polymers
(rubber), with main- or side-chain mesogens, that can {\em
  spontaneously} develop nematic orientational order. Even in the
absence of fluctuations, they were predicted\cite{GL} and later
observed to display an array of fascinating
phenomena\cite{review_theory}, the most striking of which is the
vanishing of stress for a range of strain, applied transversely to the
nematic direction. This striking softness is generic, stemming from
the spontaneous orientational symmetry breaking by the nematic
state\cite{GL,elast_us} that ensures the presence of a zero-energy
Goldstone mode, corresponding to the observed\cite{stress_strainExps}
soft distortion and strain-induced director reorientation.  The hidden
rotational symmetry also guarantees a vanishing of one of the five
elastic constants\cite{elast_us} that usually characterize harmonic
deformations of a three-dimensional uniaxial
solid\cite{LandauLifshitz}. Thermal fluctuations lead to
Grinstein-Pelcovits-like\cite{GrinPel} renormalization of elastic
constants\cite{GL} in bulk systems with dimensions below $3$ in pure
systems\cite{Xing-Radz1,StenLub} and below $5$ when effects of the
random network heterogeneity is taken into account\cite{Xing-Radz2}.
It is, therefore, likely, and indeed we find, that the elastic
properties of a two-dimensional fluctuating tensionless sheet of such
nematic elastomer differ {\em qualitatively} from those of previously
studied crystalline membranes\cite{JWS}. Our aim here is to explore
the effects of thermal fluctuations on this new universality class of
solid membranes.

Our other motivation for exploring the physics of nematic
elastomer membranes comes from earlier discovery by Toner and one
of us\cite{RTtubule}, that any amount of any kind of in-plane
anisotropy, a seemingly innocuous generalization, significantly
enriches the phase diagram of polymerized membranes.  Most
dramatically, it was predicted\cite{RTtubule} that an entire new
phase of membranes, called the ``tubule'' phase, always intervenes
between the high-temperature crumpled and low-temperature ``flat''
phases.  The defining property of the tubule phase is that it is
crumpled in one of the two membranes directions but ``flat'' (i.e.,
extended) in the other, with its thickness, roughness, and
anomalous elasticity displaying universal behaviors
controlled by a nontrivial, infrared stable fixed
point\cite{RTtubule}.  The tubule phase has since been observed in
Monte Carlo simulations\cite{BFT} of non-self-avoiding (i.e.,
phantom) anisotropic membranes with properties (the thickness,
roughness, and Poisson ratio) in close qualitative and
quantitative accord with the predictions of Radzihovsky and
Toner\cite{RTtubule}.

Although it was an {\em explicit} rotational symmetry
breaking (anisotropy) that was considered by the authors of
Ref.\ \onlinecite{RTtubule}, we expect a similar, but
qualitatively distinct tubule phase to be also displayed by
the {\em spontaneously} anisotropic elastomer membranes
considered in this paper. However, we leave the subject of
the global phase diagram, the elastomer tubule phase, and the
phase transitions between it and the flat and crumpled phases
for a future publication\cite{tubule_unpublished}. Here we
instead focus on the formulation of models of
nematically-ordered elastomer membranes and the study of
long-length-scale properties of its flat phase.

Past extensive investigations, both in the context of defected
crystalline membranes\cite{RadzihovskyNelson,MorseLubensky} and liquid
crystals confined in rigid gels (e.g., aerogels and
aerosils)\cite{RTaerogel}, demonstrated that arbitrarily weak
heterogeneity qualitatively modifies the long-scale nature of
liquid-crystal orders, membranes morphologies and elasticities.  We,
therefore, expect that local heterogeneity in an elastomer network
will also become qualitatively important on sufficiently long scales
and will likely dominate over the thermal fluctuation effects that are
the subject of this paper. However, here we consider the idealized
limit of elastomer membranes in which the effective quenched disorder
is sufficiently weak that its effects are unimportant on scales
shorter than some very long disorder-determined length scale. We leave
the study of these heterogeneity effects for a future
investigation\cite{future_disorder}.

This paper is organized as follows.  In Sec.\ \ref{model}, we develop
two models for $D$-dimensional nematically-ordered flat phase of
elastomer membranes fluctuating in $d$-dimensions with $d>D$, one with
uniaxial nematic order, the other with $D$-axial nematic order. They
are referred as uniaxial model and $D$-axial model respectively in
this paper. For the physical case of interest $D=2$, they reduce to
the same model, which we study in more detail in Sec.\ 
\ref{sec:2Dmem}.
In Sec.\ \ref{sec:harmonic}, we investigate harmonic
fluctuations of
both models and show, in particular, that in-plane phonon fluctuations
in $D\geq 3$ have parts that are equivalent to those of smectic liquid
crystal, a columnar liquid crystal, and a crystalline solid. In Sec.\ 
\ref{sec:2Dmem}, we investigate mean-field theory and harmonic
fluctuations in physically realizable two-dimensional membranes
embedded in a three-dimensional space. We find that harmonic
fluctuations in these membranes do not destroy long-range in-plane
nematic order in spite of enhanced fluctuations relative to an
isotropic system arising from the soft Goldstone mode. In Sec.
\ref{AnharmonicModel} we consider the effects of anharmonicities and
develop an $\epsilon= (4-D)$-expansion about the upper critical
dimension, $D_c=4$, for two model systems that have well-defined
analytic continuations to $D=2$.  In both models, we find similar flow
equation for coupling constants and an infrared-stable fixed point in
which the bending modulus $\kappa$ is unrenmormalized and in-plane
elastic moduli $\lambda_{ij}$ vanish with wavenumber $q$ as
$q^{\epsilon}$.

\section{Models}
\label{model}

\subsection{Notation: Reference and Target Spaces}
Although physical membranes are two-dimensional manifolds in a
three-dimensional space, it is often useful to consider
generalizations to $D$-dimensional manifolds embedded in a
$d$-dimensional space with $d>D$.  Our interest is in solid
membranes, which, unlike fluid membranes, have a non-vanishing
shear modulus at least in their isotropic state.

To describe the geometry and fluctuations of these membranes, we need
to introduce a certain amount of notation. First, we define the
reference space $S_R$.  This is the $D$-dimensional space occupied by
the membrane in its quiescent {\em flat} reference state.  We denote
$D$-dimensional vectors in this space in bold and their components
with Roman subscripts $i,j = 1, ..., D$ . In particular, we denote
intrinsic coordinates on the membrane by the vector
\begin{equation}
\rv = (x_1, ... , x_D)
\end{equation}
with components $x_i$.  The positions, $\rv$ index mass points in a
Lagrangian description of the membrane. The membrane fluctuates in a
$d$-dimensional target space $S_T$.  We denote vectors in the
embedding space with arrows and components of these vectors with Greek
subscripts $\mu, \nu = 1, ..., d$.  In particular, we describe the
position in $S_T$ of the mass point labeled by $\rv$ with the
embedding vector
\begin{equation}
\Rv(\rv) = (R_1(\rv) , ... , R_d (\rv))
\end{equation}
with components $R_{\mu} ( \rv )$.

The reference space $S_R$ is identified as a subspace of $S_T$. The
use of orthonormal basis vectors will simplify some of our discussion,
and we introduce $d$ vectors $\ev_{\mu}$ in $S_T$ with components
$\ev_{\mu \nu}= \delta_{\mu\nu}$ satisfying
\begin{equation}
\ev_{\mu} \cdot \ev_{\nu} = \delta_{\mu \nu}.
\end{equation}
Any vector in $S_T$ can be decomposed into its components along
these vectors. For example $\Rv = R_{\mu} \ev_{\mu}$, where the
summation convention on repeated indices is understood. We choose
the set $\{\ev_{\mu}\}$ so that the subset of vectors $\ev_i$, $i
= 1, ... , D$ lie in $S_R$, which we represent as a subspace of
$S_T$. Thus
\begin{eqnarray}
\ev_i \cdot \ev_i & = & \delta_{ij} \\
\ev_{\mu}\cdot \ev_{i} &= &\delta_{\mu i} =
\begin{cases}
1 & \text{if $i=\mu\leq D$} \\
0 &\text{if $\mu = D+1, ..., d$.}
\end{cases}
\end{eqnarray}
We choose length scales such that the position in $S_T$ of the point
$\rv$ in the reference membrane is
\begin{equation}
\Rv_{\rm ref} = x_i \ev_i ,
\end{equation} 
describing a perfectly flat membrane configuration.  Positions $\Rv(
\rv )$ of the distorted membrane can be described by a $D$-dimensional
in-plane phonon field $\uv(\rv)$ and a $d-D\equiv d_c$-dimensional
out-plane undulation (height) field $\hv(\rv)$:
\begin{equation}
\Rv(\rv) =  \sum_{i=1}^D (x_i + u_i(\rv))\ev_i +
 \sum_{k=1}^{d_c} h_k(\rv)\ev_{k+D} \label{Rv} .
\end{equation}

Local distortions of the membrane can be expressed in terms of the
$d\times D$-dimensional deformation tensor $\mm{\Lambda}$
\footnote{It  has to satisfy certain condition of integrability.
We will ignore this subtlety in the remaining of this paper.}
with components
\begin{equation}
\Lambda_{\mu i} = {\partial R_{\mu} \over \partial x_i} \equiv
\partial_i R_{\mu} .
\end{equation}
In the flat reference state,
\begin{equation}
\Lambda_{\mu i}^{\rm ref}={\partial R_{{\rm ref},\mu}\over\partial
x_i} = \delta_{\mu i} .
\end{equation}
Under independent rotations in the target and reference spaces,
$\Lambda_{\mu i}$ transforms as
\begin{equation}
\Lambda_{\mu i} \rightarrow \O_{T \mu \nu}\Lambda_{\nu j} \O_{R
ji}^{-1}
\label{Lambda-trans}
\end{equation}
where $\O_{T \mu \nu}$ is a $d\times d$ rotation matrix in $S_T$
and $\O_{R ij}$ a $D\times D$ rotation matrix in $S_R$.

Since $\Rv$ is a Euclidean vector, the metric tensor, which
measures distances between neighboring points via $dR^2 = g_{ij}
dx_i dx_j$, for a membrane is
\begin{equation}
g_{ij} = \frac{\partial \Rv(\rv)}{\partial x_i}\cdot
\frac{\partial \Rv(\rv)}{\partial x_j}.
\end{equation}
In the reference flat phase, $g_{ij} = g_{ij}^{\rm ref} = \Lambda_{\mu
  i}^{\rm ref} \Lambda_{\mu j}^{\rm ref}= \delta_{ij}$. We will denote
tensors in $S_R$ with a double underscore.  Thus the metric tensor is
$\mm{g}$. The Lagrangian strain tensor $\mm{u}$, with components
$u_{ij}= g_{ij} - g_{ij}^{\rm ref}$, measures the local change in
separations of nearby points in distorted states characterized by
$\Rv(\rv)$ relative to those of the undistorted, flat state
characterized by $\Rv_{\rm ref}$:
\begin{equation}
dR^2 - dR_{\rm ref}^2 = 2 u_{ij} dx_i dx_j .
\label{length}
\end{equation}
Using Eq.\ (\ref{Rv}), the strain can be expressed in terms of the
phonon and height fields as
\begin{eqnarray}
u_{ij}&=& {1 \over 2} ( g_{ij}-g^{\rm ref}_{ij})= {1\over
2}\left(\frac{\partial \Rv}{\partial x_i}\cdot \frac{\partial
\Rv}{\partial x_j}
-\delta_{ij}\right) \nonumber\\
&=& \case{1}{2}(\partial_i u_j + \partial_j u_i +
\partial_i \uv \cdot \partial_j \uv + \partial_i \hv\cdot \partial_j
\hv).
\label{uij}
\end{eqnarray}

\subsection{The Model Hamiltonian}
As discussed in the Introduction, internally isotropic solid
membranes have been extensively studied\cite{JWS}. In contrast to
liquid membranes and one-dimensional polymer analogs, they admit a
flat phase, characterized by an infinite orientational (in
embedding space) persistence length that is stable to thermal
fluctuations.  The long-wavelength free-energy density for
distortions of such spontaneously flat phase is given by
\begin{equation}
{\calF}_I = {\calF}_{{\rm bend}} + {\calF}_{{\rm stretch}} ,
\label{H_I}
\end{equation}
where
\begin{equation}
{\calF}_{{\rm bend}} = \frac{\kappa}{2}(\nabla ^2\hv)^2
\label{H_bend}
\end{equation}
and
\begin{subequations}
\label{H_S}
\begin{eqnarray}
{\calF}_{{\rm stretch}} & = &
  \frac{\lambda}{2} (\Tr\um)^2+\mu(\Tr\um^2)
\label{H_Ia}\\
&=& \frac{B}{2}(\Tr\um)^2+\mu(\Tr\tilde{\um}^2),
\label{H_Ib}
\end{eqnarray}
\end{subequations}
where $\kappa$ is membrane's bending
rigidity\cite{tensionless}, $\lambda$ and $\mu$ are the
Lam\'e coefficients characterizing in-plane elasticity,
$B=\lambda + 2\mu/D$ is the bulk elastic modulus, and
\begin{equation}
\tilde{u}_{ij}=u_{ij}-\frac{1}{D}(\Tr\um)\delta_{ij}
\end{equation}
is the symmetric-traceless part of the strain tensor with
$\Tr\tilde{\um}=0$. Earlier studies of non-liquid crystalline
polymerized membranes\cite{NP,AGL,LR} have demonstrated that thermal
fluctuations produce wild undulations about the flat state $\Rv_{ref}$
that make elastic nonlinearities in the height field $\hv$ (but not in
the in-plane phonon fields $\uv$) in $u_{ij}$ Eq.\ (\ref{uij})
important on long length scales and lead to universal
length-scale-dependent elastic moduli: $\kappa(\qv) \sim
q^{-\eta_{\kappa}}$, $\lambda(\qv)\sim q^{\eta_\mu}$, and
$\mu(\qv)\sim q^{\eta_\mu}$. One of the most important consequences of
this is that the thermally-driven upward renormalization of the
bending rigidity $\kappa$ stabilizes the low-temperature flat phase of
two dimensional \cite{MerminWagner} membranes against these very same
fluctuations\cite{NP,AGL,LR}.

The free energy density of Eq.\ (\ref{H_I}) provides a correct
description of elastic and height fluctuations of a membrane so long
as the equilibrium phase is truly an isotropic flat phase.  If,
however, the shear modulus $\mu$ becomes sufficiently small, the
membrane becomes unstable to spontaneous in-plane distortions that
break rotational symmetry of
$S_R$\cite{GL,elast_us,tubule_unpublished}.  In the presence of such
instability the distortion is stabilized by a non-linear strain energy
$\calF_{NL}$ and an in-plane curvature energy $\calF_{{\rm curv}}$,
neglected in $\calF_I$, Eqs.\ref{H_I}-\ref{H_Ib}.  The in-plane
curvature energy, which stabilizes the system against spatially
nonuniform distortions is simply
\begin{equation}
\calF_{{\rm curv}} = \case{1}{2} K (\nabla^2 \uv )^2 .
\end{equation}
The important low-order contributions to non-linear strain energy
are
\begin{equation}
\calF_{NL} = \beta\Tr\um\Tr\tilde{\um}^2 -C \Tr (\tilde{\um} )^3+
\gamma(\Tr \tilde{\um}^2)^2 .
\label{H_NL}
\end{equation}
Terms of order $\um^5$ or higher and terms proportional to $(\Tr \um
)^3$ and $(\Tr \um)^2 \Tr \tilde{\um}^2$ could also be added but are
qualitatively unessential for our current discussion.

\subsection{Spontaneously Anisotropic Phases}
When $\mu$ becomes sufficiently small in the presence of
$\calF_{NL}$, a membrane will undergo a
transition\cite{GL,elast_us,tubule_unpublished}
to a new anisotropic, spontaneously stretched
equilibrium state with a non-vanishing equilibrium strain
$\mm{u}_0$, an anisotropic deformation tensor $\Lambda_{\mu i}^0$
that differs from $\Lambda_{\mu i}^{\rm ref}=\delta_{\mu i}$, and
position vectors $R_{0,\mu} = \Lambda_{\mu i}^0 x_i$. The most
general form of $\Lambda_{\mu i}^0$ can be obtained via target-
and reference-space rotations Eq.\ (\ref{Lambda-trans}) of the
deformation tensor
\begin{equation}
\Lambda_{\mu j}^0 = \begin{cases}
\Lambda_{ij}^0 & \text{if $\mu = i = 1, ... ,D$} \\
0 & \text{if $\mu = D+ 1, ... ,d$ ,}
\end{cases}
\end{equation}
resulting from a distortion without rotation in the $S_R$ subspace
of $S_T$. Here $\Lambda_{ij}^0$ are the components of the a $D
\times D$ matrix $\mm{\Lambda}^0$ restricted to the plane of the
original reference space $S_R$. The new equilibrium
$\mm{\Lambda}^0$ yields a new metric tensor $g_{ij}^0 =
\Lambda_{\mu i}^0 \Lambda_{\mu j}^0$ and a non-vanishing
equilibrium strain relative to the original isotropic flat
membrane:
\begin{equation}
u_{ij}^0 = {1 \over 2} \left( \Lambda_{\mu i}^0 \Lambda_{\mu j}^0
- \delta_{ij} \right).
\label{equil-nem-strain}
\end{equation}

The simplest anisotropic state that can form is the uniaxial one in
which the original isotropic membrane is stretched (compressed) along
a single direction in $S_R$ and compressed (stretched) along the
others. Because the reference state is isotropic, the direction of
stretching, specified by a unit vector $\nv^0$ with components
$n_i^0$, is arbitrary in the plane $S_R$, and we could take it without
loss of generality to be along one of the basis vectors, say $\ev_1$.
The vector $\nv^0$ also exists in the target space $S_T$.  It has
components $n_{\mu}^0$ that are equal to $n_i^0$ for $\mu = i = 1, ...
, D$ and zero for $\mu > D$. Thus, in the uniaxial case,
\begin{equation}
\Lambda_{ij}^0 = (\Lambda_{0||} - \Lambda_{0\perp} ) n_{0i}
n_{0j} + \Lambda_{0\perp} \delta_{ij} .
\end{equation}
Under rotations in $S_T$ and $S_R$, this becomes
\begin{equation}
\Lambda_{\mu i}^0 =(\Lambda_{0||} - \Lambda_{0\perp} )n_{\mu}^T
n_i^R + \Lambda_{0\perp} \ev_{\mu}^T \cdot\ev_i^R ,
\end{equation}
where we used $\delta_{ij} =\ev_{\nu i} \ev_{\nu j}$ and where
$n_{\mu}^T = \O_{T\mu \nu} n_{0\nu}$, $\ev_{\nu \mu}^T = \O_{T \mu
i}\ev_{\nu i}$, $\ev_{\nu i}^R = \O_{ij}^R \ev_{\nu j}$, and
$n_i^R = \O_{ij}^R n_{0j}$.

The opposite limit of the uniaxial anisotropic state is the full
$D$-axial state in which there is unequal stretching in all $D$
directions in the plane.  In this case,
\begin{equation}
\Lambda_{ij}^0 = \sum_{k=1}^D \Lambda_{0k} e_{k i} e_{k j}
\end{equation}
with all $\Lambda_{0k}$ different.

As discussed in detail in Ref. \cite{elast_us}, the spontaneous
broken symmetry described by the above deformation tensors leads
to vanishing of certain shear moduli of the distorted solid.
Distortions relative to the new anisotropic state are measured by
the displacement and height vectors, $\uv'(\rv')$ and ${\vec h} (
\rv' )$ defined via
\begin{equation}
\Rv'(\rv' ) = \rv' + \uv'( \rv' ) + {\vec h} ( \rv')\equiv \Rv(
\rv) ,
\end{equation}
where by definition, $\rv' = \Rv_0$, $\uv'(\rv')$ is the the
$D$-dimensional in-plane displacement vector, and $\hv(\rv)$ is
the $d_c$-dimensional ($d_c = D-d$) height variable orthogonal to
$\uv'$. The strain tensor relative to the new state,
\begin{equation}
u'_{ij} = {1\over 2} \left( {\partial R_{\mu} \over \partial
x_i'}{\partial R_{\mu} \over \partial x_j'} - \delta_{ij} \right)
\end{equation}
expressed in terms of the strain tensor $\mm{u}$ relative to the
original state is
\begin{equation}
\um = \um_0 + \Lm^T_0 \mm{u'}\Lm_0 .
\label{u-u'}
\end{equation}
The strain Hamiltonian density to harmonic order for the uniaxial
case
\begin{equation}
\calF^{{\rm uni}}_{\rm strain}=\case{1}{2}\sum_{ij}\lambda^{\rm
uni}_{ij}u'_{ii} u'_{jj} + \mu_{\perp} u_{ij}^{\prime\perp}
u_{ij}^{\prime\perp}
\label{nem-energy}
\end{equation}
where the Einstein convention is not used in the first term on the
right hand side (but is in the second),
\begin{equation}
u_{ij}^{\prime\perp} = \delta_{ik}^Tu'_{kl} \delta_{lj}^T ,
\end{equation}
\begin{equation}
\delta_{ij}^T = \delta_{ij} - n_{0i} n_{0j} ,
\end{equation}
and
\begin{eqnarray}
\lambda^{\rm uni}_{ij} &=&\lambda_1 \delta_{i1}\delta_{j1} +
\lambda_3 (1-
\delta_{i1})(1- \delta_{j1})\nonumber \\
& & + \lambda_2[\delta_{i1}(1- \delta_{j1}) + \delta_{j1}(1-
\delta_{i1})] ,
\label{lambda-uni}
\end{eqnarray}
where the values of $\lambda_1$, $\lambda_2$ and $\lambda_3$
depend on the potentials of the original Hamiltonian and
$\mm{\Lambda_0}$. This form makes it clear that
$\lambda_{ij}$ is actually a subset of a fourth rank tensor
and not a true second-rank tensor.
 The bending and curvature energies become
anisotropic with
\begin{eqnarray}
\calF_{{\rm bend}} &= &\case{1}{2} \kappa_{ij}
\partial_i^{\prime 2}
{\vec h} \cdot \partial_j^{\prime 2} {\vec h} \\
\calF_{{\rm curv}}& = &\case{1}{2} K_{ij} \partial_i^{\prime 2}
{\uv'}\cdot \partial_j^{\prime 2} {\uv'}
\label{bend_curv}
\end{eqnarray}
where the anisotropic bending and curvature moduli can be
expressed, respectively, as $\kappa_{ij} = \kappa
A^{{\rm uni}}_{ij}$ and $K_{ij} = K A^{{\rm uni}}_{ij}$ where
\begin{eqnarray}
A^{{\rm uni}}_{ij}&  = &\Lambda_{0||}^4 \delta_{i1}\delta_{j1} +
\Lambda_{0\perp}^4 (1- \delta_{i1}) (1-\delta_{j1})  \\
& & + \Lambda_{0||}^2 \Lambda_{0\perp}^2 [\delta_{i1}(1-
\delta_{j1}) + \delta_{j1}(1- \delta_{i1})] .\nonumber
\end{eqnarray}
In the full $D$-axial case, there is no residual plane that can
support shears and the harmonic free-energy density depends only
on the diagonal parts of the strain:
\begin{equation}
\calF^{{\rm D-axial}}_{\rm strain} = \case{1}{2}\sum_{ij}\lambda_{ij} u'_{ii}
u'_{jj}
\label{D-axial-energy}
\end{equation}
with $\lambda_{ij}$ a symmetric matrix with all $D(D+1)/2$ independent
entries different.  Again, the Einstein convention is suspended in
this equation. The bend and curvature energies have the same form as
Eqs.\ (\ref{bend_curv}) but $A_{ij}^{{\rm uni}}$ is replaced by
\begin{equation}
A^{{\rm D-axial}}_{ij} = \Lambda_{0i}^2 \Lambda_{0j}^2.
\end{equation}
We note that in the most general case that is compatible with
D-axial symmetry, all $D(D+1)/2$ components of 
$A^{\rm D-axial}_{ij}$ are independent of each other. 

The total Hamiltonian density for a nematic elastomeric membrane is
\begin{equation}
\calF = \calF_{\rm strain} + \calF_{\rm bend} + \calF_{\rm curv}
\label{Ftot} .
\end{equation}
We have expressed $\calF_{\rm strain}$ only to harmonic order in the
nonlinear strain $u_{ij}$ even though nonlinear terms in the original
isotropic energy are essential to stabilize the system after nematic
order develops\cite{elast_us}.  As we shall see in Sec.\ 
\ref{AnharmonicModel}, near four dimensions anharmonic terms in the
nonlinear strains associated with in-plane phonons are subdominant
(irrelevant in RG sense) to nonlinearities in the height undulations
$\hv$ and we will ignore them in what follows.  This is in strong
contrast to bulk nematic elastomers, where such phonon nonlinearities
cannot be ignored and must be treated
non-perturbatively\cite{Xing-Radz1,StenLub}.

In two dimensions, which we consider in more detail in the Sec.\ 
\ref{sec:2Dmem}, there is only one direction, say the $y$ direction,
perpendicular to the direction of order, and $u^{\prime \perp}_{ij} =
\delta_{iy} \delta_{jy} u'_{yy}$.  Thus the second term in Eq.\ 
(\ref{nem-energy}) can be absorbed into the first term, which depends
only on diagonal components of $u_{ij}$. Thus, in $2D$, the free
energy density has the same form as Eq.\ (\ref{D-axial-energy}), and
is most naturally analytically continued from $D$-axial generalization
of a nematic elastomer membrane.

\section{Harmonic theory}
\label{sec:harmonic}
The two (uniaxial and D-axial) nematic membrane Hamiltonians
introduced in the preceding section has complicated anisotropies,
associated with spontaneous in-plane nematic order, that lead to
membrane's highly anisotropic conformational correlations.  In this
section, we will investigate fluctuations in the harmonic
approximation to these models.  These harmonic results will be
necessary in our subsequent discussions in Sec.\ \ref{AnharmonicModel}
of the important effects of nonlinearities in the presence of
fluctuations. To simplify the notation, in this section we will drop
primes appearing in the strain and phonon fields $u_{ij}'$ and $\uv'$
measured relative to the nematic state and denote them by $u_{ij}$ and
$\uv$, not to be confused with physically distinct unprimed quantities
of the previous section.

The harmonic Hamiltonians for the uniaxial and $D$-axial models have a
longitudinal-strain and height elastic parts describing the cost of
simple shear modes:
\begin{eqnarray}
\calH_L^0 &= & {1 \over 2}\int{d^D q\over (2
\pi)^D}\big[\sum_{i,j=1}^D[\lambda_{ij} q_i q_j +
\delta_{ij}K({\hat
q}) q^4]u_i u_j \nonumber\\
& &+ \kappa({\hat q}) q^4 \sum_{i=D+1}^d h_i h_i \big]
\label{H-harmonic}
\end{eqnarray}
where
\begin{eqnarray}
K({\hat q}) &= &\sum_{ij} K_{ij} {\hat q}_i^2 {\hat q}_j^2 \\
\kappa({\hat q}) & = & \sum_{ij} \kappa_{ij} {\hat q}_i^2 {\hat
q}_j^2
\end{eqnarray}
with ${\hat q}_i = q_i/q$. In uniaxial systems, there is in
addition a shear part in the $(D-1)$ dimensional
plane perpendicular to the unique direction $\nv_0$.
\begin{eqnarray}
\calH_T^0&=& \mu_{\perp} \int d^D x
u_{ij}^{\perp}u_{ij}^{\perp}  \\
& & \rightarrow {1 \over 2}\mu_{\perp} \int{d^D q\over (2 \pi)^D}[
q_{\perp}^2 u_i^{\perp}u_i^{\perp} + q_{\perp i} q_{\perp j}
u_i^{\perp} u_j^{\perp} ],\nonumber
\end{eqnarray}
where $\uv^{\perp}$ is the part of the vector $\uv$ in the plane
perpendicular to $\nv_0$.  $\uv^{\perp}$ has $D-1$ components. Thus in
$D=2$, $\uv^{\perp}$ has only one component, say $u_y$,
$u_{ij}^{\perp} = u_{yy}$ and two models are equivalent.

In both models, at harmonic level, height and phonon variables
decouple, and the height correlation function is simply
\begin{equation}
G_{h_i h_j} = \delta^P_{ij} {1 \over \kappa({\hat q}) q^4} ,
\label{eq:height}
\end{equation}
where $\delta^P_{ij}$ is the projection operator onto the $d_c$
dimensional subspace perpendicular to the reference membrane.

\subsection{Uniaxial Case}
In the uniaxial case, ${\bf u}$ can be decomposed into components
along the uniaxial direction $\nv_0$ and along directions
parallel and transverse to the wave-vector $\qv_{\perp}$ in
the plane perpendicular to $\nv_0$:
\begin{equation}
u_i = u_n n_{0i} + u_L {\hat q}_{\perp i} + u_{t i} ,
\label{u-rep}
\end{equation}
where $\nv_0 \cdot \uv_t = 0$ and $\qv_{\perp} \cdot \uv_t = 0$.
We will represent this decomposition as $u_i = (u_n, u_l ,\uv_t)$.
The $\uv$-correlation function can then be decomposed as
\begin{eqnarray}
G_{ij} &=& G_{nn} n_{0i} n_{0j} + G_{LL}{\hat q}_{\perp i} {\hat
q}_{\perp j} + G_{nL} (n_{0i} {\hat q}_{\perp j} + n_{0j} {\hat
q}_{\perp i})\nonumber \\
& & + G_t(\delta^{T}_{ij} - {\hat q}_{\perp i} {\hat q}_{\perp
j} ) .
\end{eqnarray}
The transverse part of $G_{ij}$ decouples from the others and is
easily calculated
\begin{equation}
G_t = {1 \over \mu_{\perp} q_{\perp}^2} .
\end{equation}
The Hamiltonian for the $u_n$ and $u_L$ components can be written
as
\begin{equation}
\calH = {1 \over 2} \int {d^D q\over (2 \pi)^D} {\tilde u}_a
G^{-1}_{ab} {\tilde u}_b ,
\end{equation}
where ${\tilde u} = (u_n, u_L)$ and
\begin{equation}
G^{-1} = \left(
\begin{array}{cc}
\lambda_1 q_{||}^2 +K({\hat q}) q^4 & \lambda_2
q_{||} q_{\perp} \\
\lambda_2 q_{||} q_{\perp} & {\bar \lambda}_{3}
q_{\perp}^2 + K({\hat q}) q^4\\
\end{array}
\right) ,
\label{eq:G-1}
\end{equation}
where ${\bar \lambda}_3 = \lambda_3+ \mu_{\perp}$ with 
$\Lambda_i$ defined in Eq.\,(\ref{lambda-uni}). In the long-wavelength
limit, terms of order $q_{||}^4$ and $q_{||}^2 q_{\perp}^2$ can be
neglected relative to $q_{||}^2$ , and the $nn$ entry in $G^{-1}$
is effectively $\lambda_1q_{||}^2 + K_{\perp}q_{\perp}^4$, which
is the inverse of the propagator for fluctuations in a smectic
liquid crystal.  Similarly, in the long-wavelength limit, the $LL$
part of $G^{-1}$ is effectively ${\bar \lambda}_3 q_{\perp}^2 +
K_{||}q_{||}^4$, which is the inverse propagator for one component
of displacement fluctuations in a columnar liquid crystal. Thus,
when $\lambda_2 = 0$, $G^{-1}$ decomposes into two independent
parts, one of which is identical to the inverse propagator of a
smectic liquid crystal and the other of which is identical to one
component of the propagator of a columnar liquid crystal.  We will
find below that this property survives the turning on of
$\lambda_2$ though with different coefficients.

Equation (\ref{eq:G-1}) is easily inverted to yield
\begin{equation}
G = {1 \over \Delta } \left(
\begin{array}{cc}
{\bar \lambda}_3 q_{\perp}^2 + K({\hat q}) q^4 & - \lambda_2
q_{||} q_{\perp} \\
- \lambda_2q_{||} q_{\perp} & \lambda_1 q_{||}^2 + K({\hat q})q^4
\end{array}
\right) ,
\label{eq:G}
\end{equation}
where
\begin{equation}
\Delta = \det G^{-1} = \Delta_{\bar \lambda} q_{\perp}^2 q_{||}^2
+ (\lambda_1 q_{||}^2 + {\bar \lambda}_3 q_{\perp}^2 ) K({\hat q})q^4 +
\cdots
\end{equation}
with $\Delta_{\bar\lambda} = \lambda_1 \bar{\lambda}_3 - \lambda_2^2$. 
It is can be shown (See Appendix A for discussion in two dimension)
that $G_{nn}$ is dominated at small $\qv$ by the region with $q_{||}
\sim q_{\perp}^2$ so that
\begin{equation}
G_{nn} = {{\bar \lambda}_3 q_{\perp}^2 + K({\hat q})q^4 \over
\Delta} \rightarrow {1 \over \lambda_1'q_{||}^2 + K_{\perp}
q_{\perp}^4} ,
\end{equation}
where $\lambda_1' = \Delta_{\bar \lambda}/{\bar\lambda}_3 =
\lambda_1 - \lambda_2^2/{\bar\lambda}_3$, $K_{\perp}
= L (\Lambda_{0\perp})^4$. Similarly, $G_{LL}$ is
dominated by $q_{\perp}\approx q_{||}^2$ at small $\qv$ and
\begin{equation}
G_{LL} ={\lambda_1 q_{||}^2 + K({\hat q})q^4 \over \Delta}
\rightarrow {1 \over \lambda_3'q_{\perp}^2 + K_{||} q_{||}^4} ,
\end{equation}
where $\lambda_3' = \Delta_{\bar \lambda}/\lambda_1$,
$K_{||} = K (\Lambda_{0||})^4$.

Thus we find, that, within the harmonic approximation, the
translational lower critical dimension is determined by the dominant
smectic-like fluctuations of $u_n$ modes and is equal by $D=3$. The
subdominant columnar-like modes $u_L$ become important only below
$D=5/2$, while the transverse fluctuations $\uv_t$ only grow with
length scale for $D \leq 2$.

\subsection{$D$-axial systems}
In $D$-axial systems, the strain energy is given by Eq.\
(\ref{D-axial-energy}) with $D(D+1)/2$ independent components for
$\lambda_{ij}$. The harmonic Hamiltonian is then given by Eq.\
(\ref{H-harmonic}) with no transverse contributions.  The height
fluctuations are the same as in the uniaxial case with the
appropriate expression for $\kappa ({\hat q})$.  The inverse
phonon propagator is
\begin{equation}
G^{-1} = \begin{pmatrix}
\lambda_1 q_1^2 + K({\hat q})q^4 & \ldots &
    \lambda_{1D} q_1 q_D \\
    \vdots & \ddots & \vdots \\
    \lambda_{1D} q_1 q_D& \ldots & \lambda_D q_D^2 + K({\hat q}) q^4
\end{pmatrix}
.
\label{eq:D-axialG-1}
\end{equation}
$G$ can be obtained by inverting $G^{-1}$ using minors: $G_{ij} =
(-1)^{i+j}{\tilde \Delta}^{ji}/\Delta$ where $\Delta= \det G^{-1}$
and ${\tilde \Delta}^{ij}$ is the determinant of the minor of
$G^{-1}$ obtained by deleting row $i$ and column $j$.  The results
for the long-wavelength diagonal and off-diagonal components of
$G_{ij}$ are, respectively,
\begin{subequations}
\label{eq:D-axialG}
\begin{eqnarray}
G_{ii} & \approx & {1 \over \lambda'_{ii} q_i^2 + K({\hat q})q^4}, \\
G_{ij} & \approx & {(-1)^{i+j} {\tilde \Delta}_{\lambda}^{ji} q_i q_j
\over \Delta_{\lambda}q_i^2 q_j^2 + K({\hat q}) q^4 ( {\tilde
\Delta}_{\lambda}^{ii} q_j^2 + \Delta_{\lambda}^{jj} q_i^2 )} ,\ \ \ \ \
\end{eqnarray}
\end{subequations}
where, as before, $\Delta_{\lambda}$ is the determinant of $\lambda$,
${\tilde \Delta}_{\lambda}^{ij}$ is the determinant of the $ij$-minor
of $\lambda$, and $\lambda'_{ii} = \det\Delta_{\lambda}/{\tilde
\Delta}_{\lambda}^{ij}$. Thus, the diagonal components $G_{ii}$ have
the structure of a smectic propagator, with all terms involving $q_i$
in $K({\hat q}) q^4$ subdominant in the long wavelength limit.

\section{Two-Dimensional Membranes}
\label{sec:2Dmem}

As discussed in the Introduction, real membranes are two-dimensional
manifolds, with reference-space coordinates $\rv = (x_1, x_2)$,
fluctuating in a three-dimensional space with coordinates ${\vec R} =
(R_1, R_2, R_3)$.  Because of their low dimensionality, these
membranes have a number of special properties, some of which we will
investigate in this section.  In particular, we will study the
transition from an isotropic to a nematically-ordered membrane, which,
because of the absence of a third-order invariant for two-dimensional
symmetric-traceless tensors is of second order in mean-field
theory. Our treatment will review in a two-dimensional context the
mechanism for the emergence of soft elasticity in the nematic phase.
We will also consider harmonic fluctuations about the flat nematic
state in some detail and show that they lead to short-ranged
positional correlations, but allow for a stable long-range
orientational (nematic) in-plane order of the membrane.

\subsection{The Two-Dimensional Hamiltonian}
Our starting point is the elastic energy of an isotropic
two-dimensional membrane augmented by nonlinear elastic terms
necessary to stabilize the nematic state that develops when the shear
modulus becomes negative.  A simplifying feature special to two
dimensions is that the strain tensor $\mm{u}$ is a $2\times 2$ matrix
with no independent cubic invariant. The corresponding Hamiltonian is
\begin{subequations}
\label{Heffb}
\begin{eqnarray}
{\calF} &=&\frac{\kappa}{2} (\nabla^2 \hv)^2 +\frac{K}{2}
(\nabla^2 \uv)^2
 + \frac{B}{2}(\Tr\um)^2 + \mu\Tr\tilde{\um}^2\nonumber\\
&+& \beta\Tr\um\Tr\tilde{\um}^2 + \gamma(\Tr \tilde{\um}^2)^2
\label{Heffa}\\
&=&\frac{\kappa}{2}(\nabla^2 \hv)^2 + \frac{K}{2}(\nabla^2 \uv)^2 +
 \frac{B}{2}(\Tr\um+\frac{\beta}{B}\Tr\tilde{\um}^2)^2 \nonumber\\
 &+&(\gamma-\frac{\beta^2}{2 B}) \left( \Tr \tilde{\um}^2 +
 \frac{\mu/2}{\gamma-\beta^2/(2 B)} \right)^2.
\end{eqnarray}
\label{Heff}
\end{subequations}
To simplify our discussion, we have left out the cubic and quartic
terms, $(\Tr\um)^3$, $(\Tr\um)^4 $ and $(\Tr\um)^2
(\Tr\tilde{\um}^2)$, as well as higher order nonlinearities in
$\mm{u}$. The inclusion of these terms, which are quantitatively
smaller than the terms we keep in the nearly incompressible (large
$B$) limit, will not qualitatively modify our results.

The nonlinear strain tensor $u^0_{ij}$ Eq.\
(\ref{equil-nem-strain}) associated with the spontaneous
deformation in the nematic state can easily be found by
minimizing the effective energy Eq.\ (\ref{Heffb}). Since
$\Tr\tilde{\um}_0^2$ is greater than or equal to zero, its
value $\Tr\tilde{\um}_0^2$ at equilibrium is zero so long as
$\mu>0$. In this case, $\Tr\um_0$ is also zero.  When
$\mu<0$, $\Tr\tilde{\um}_0^2$ and $\Tr\um_0$ become nonzero
with values
\begin{subequations}
\label{tr_u0}
\begin{eqnarray}
\Tr\tilde{\um}_0^2 &=& \frac{B |\mu|}{2 B\gamma-\beta^2},\\
\Tr\um_0&=& -\frac{\beta}{B} \Tr \tilde{\um}_0^2\nonumber,\\
          &=& -\frac{\beta |\mu|}{2 B\gamma - \beta^2} .
\end{eqnarray}
\end{subequations}
By choosing axes so that the anisotropic stretching is along
the $x$-axis, we can express $\um_0$ and $\tilde{\um}_0$ in terms
of $\Lambda_{0x}$ and $\Lambda_{0y}$ as
\begin{subequations}
\begin{eqnarray}
\um_0& = & {1\over 2}\left(\begin{array}{cc} \Lambda_{0x}^2 - 1 &
0 \\ 0 & \Lambda_{0y}^2 - 1 \end{array} \right), \\
\tilde{\um}_0 & = & {1 \over 4} (\Lambda_{0x}^2 - \Lambda_{0y}^2 )
\left(\begin{array}{cc} 1 & 0 \\
0 & -1 \end{array} \right) .
\end{eqnarray}
\end{subequations}
 From this and
\begin{subequations}
\label{u0}
\begin{eqnarray}
\mm{u}_0 &=& \case{1}{2}(\Tr\um_0)\Im + \tilde{\um}_0 \\
    &=& \case{1}{2}(\Tr\um_0)\Im +
    \case{1}{2}\sqrt{2\Tr\tilde{\um}_0^2}\,
       \left(\begin{array}{cc} 1 & 0 \\
                                    0 &-1
\end{array}\right),
\label{eqdeform}
\end{eqnarray}
\end{subequations}
we obtain
\begin{eqnarray}
\Lambda_{0x} & = & \left(1 + \Tr\tilde{\um}_0 + \sqrt{2
\Tr\tilde{\um}_0^2}\right)^{\frac{1}{2}}, \\
\Lambda_{0y} & = & \left(1 + \Tr\tilde{\um}_0 - \sqrt{2
\Tr\tilde{\um}_0^2}\right)^{\frac{1}{2}} ,
\end{eqnarray}
where ${\rm Tr}{\tilde \um}_0$ and $\Tr\tilde{\um}_0^2$ are given
in Eq.\ (\ref{tr_u0}).

We can now massage the free energy $\calF$, Eq.\ (\ref{Heffb}),
into a more convenient form,
\begin{eqnarray}
\calF
&=& \frac{\kappa}{2}(\nabla^2 h)^2 + \frac{K}{2}(\nabla^2 \uv)^2\nonumber\\
&+& \frac{B}{2}(\Tr\um-\Tr\um_0)^2
 + \gamma(\Tr\tilde{\um}^2 - \Tr\tilde{\um}_0^2)^2\nonumber\\
&+&\beta(\Tr\um-\Tr\um_0)(\Tr\tilde{\um}^2-\Tr\tilde{\um}_0^2),
\label{Hu_0}
\end{eqnarray}
that makes it transparent that $\calF$ is minimized by $\um=\um_0$
and that permits a straightforward expansion about the ground
state.

Using Eq.\ (\ref{u-u'}), we can now easily express the
free-energy density of Eq.\ (\ref{Hu_0}) in terms of the
strain $\um'$ relative to the new stretched equilibrium
state. First we observe
\begin{subequations}
\label{um-expan}
\begin{eqnarray}
\Tr \um-\Tr \um_0           &=&   \Tr \Lm_0^2 u',  \\
 \Tr \tilde{\um}^2-\Tr \tilde{\um}_0^2  &=&
2 \Tr(\tilde{\um}_0 \Lm _0^2 \um') + \Tr(\Lm _0^2 \um')^2 \nonumber\\
                   &-& \frac{1}{2} (\Tr \Lm_0^2 \um')^2 \\
                   &\approx&\case{1}{2}(\Lambda_{ox}^2 - \Lambda_{0y}^2)
                   (\Lambda_{0x}^2 u'_{xx} - \Lambda_{0y}^2 u'_{yy})
                   .\nonumber
\end{eqnarray}
\end{subequations}
The final expression, valid to linear order in $\um'$, does not
depend on $u'_{xy}$ -- a property, whose origin is the spontaneous
broken rotational symmetry of the nematic phase, that is
responsible for the vanishing of the membrane shear modulus. Using
Eq.\ (\ref{um-expan}) in Eq.\ (\ref{Hu_0}), replacing $\partial_i$
by $\partial_i'=\Lambda_{0ij} \partial_j$, and retaining only the
dominant terms in $\partial_i \uv$, we obtain
\begin{eqnarray}
{\calF} &=& \case{1}{2} \kappa_{xx}(\partial_x^2\hv)^2
  + \case{1}{2}\kappa_{yy}(\partial_y^2 \hv)^2
  + \kappa_{xy} (\partial_x^2\hv)\cdot(\partial_y^2\hv)\nonumber\\
& & +\case{1}{2}K_y (\partial_y^2 u_x)^2
   +\case{1}{2}K_x(\partial_x^2 u_y)^2  \nonumber \\
& & +\case{1}{2} \lambda_{x} u_{xx}^2 + \case{1}{2}\lambda_y
u_{yy}^2 + \lambda_{xy} u_{xx}u_{yy} ,
\end{eqnarray}
where to streamline our notation we again dropped primes, by replacing
$\um'$ with $\um$ and $\partial'_i$ with $\partial_i$ . The strains
are the usual nonlinear strains relative to the new reference state,
which to linear order are simply $u_{xx} =\partial_x u_x$ and $u_{yy}
= \partial_y u_y$. The bare elastic coefficients $\kappa$'s, $K's$,
and $\lambda$'s are determined by the parameters of the original
Hamiltonian and $\Lm_0$:
\begin{eqnarray}
\kappa_{xx}&=&\Lambda_{0x}^4 \kappa,\nonumber\\
\kappa_{xy}&=&\Lambda_{0x}^2 \Lambda_{0y}^2 \kappa,\nonumber\\
\kappa_{yy}&=&\Lambda_{0y}^4 \kappa,\nonumber\\
K_x &=& \Lambda_{0y}^4 K,\nonumber\\
K_y &=& \Lambda_{0x}^4 K,\nonumber\\
\lambda_x &=& \Lambda_{0x}^4  (B
           +  \beta (\Lambda_{0x}^2-\Lambda_{0y}^2 )
                 + \case{1}{2}\gamma (\Lambda_{0x}^2-\Lambda_{0y}^2)^2 ),
\nonumber\\
\lambda_y &=& \Lambda_{0y}^4
               (B -  \beta (\Lambda_{0x}^2- \Lambda_{0y}^2)
                 + \case{1}{2}\gamma (\Lambda_{0x}^2-\Lambda_{0y}^2)^2 ),
\nonumber\\
\lambda_{xy} &=&  \Lambda_{0x}^2 \Lambda_{0y}^2 (B
                           - \gamma (\Lambda_{0x}^2-\Lambda_{0y}^2)^2 ).
\end{eqnarray}

\subsection{Fluctuations and Correlations of the Harmonic Model}
\label{HarmonicModel}

We study fluctuations of nematically-ordered elastomer membranes
within a harmonic approximation, for the physically realizable case of
$D=2$ and $d=3$. In this case, the displacement vector $\uv$ has two
components, and the height has a single component $h$. Within this
approximation all correlation functions are related to the harmonic
two-point correlation functions\cite{CL}
\begin{subequations}
\begin{eqnarray}
{G}^{0}_{h}(\rv)&=&\langle h(\rv) h(0)\rangle_0\nonumber\\
&=&\int{d^2 q\over(2\pi)^2}\,e^{i\qv\cdot\rv} G^0_h(\qv),
\label{Gh}\\
{G}^0_{ij}(\rv)&=&\langle u_i(\rv) u_j(0)\rangle_0\nonumber\\
&=&\int{d^2 q\over(2\pi)^2}\,e^{i\qv\cdot\rv}\,G^0_{ij}(\qv),
\label{Gu}
\end{eqnarray}
\end{subequations}
expressed in terms of corresponding ``propagators''
$G^0_h(\qv)$ and $G_{ij}^0(\qv)$.  As usual, the averages are
computed using a Boltzmann weight
$Z_0^{-1}e^{-\calH_0[\hv,\uv]}$ (for convenience using $k_B
T$ as the energy unit), by integrating over phonon $\uv$ and
height undulation $\hv$ fields, with $Z_0=\int D\hv D\uv\
e^{-\calH_0}$ the partition function and $\calH_0$ the
harmonic effective Hamiltonian $\calH_0=\int d^2
x\calF_0[\hv,\uv]$ obtained from effective Hamiltonian,
$\calH=\int d^2 x\calF[\hv,\uv]$ by neglecting all elastic
nonlinearities appearing in Eq.\ (\ref{uij}).

The height propagator $G_{h}^0(\qv)$ is given by Eq.\ (\ref{eq:height})
and the phonon propagator $G_{ij}^0 ( \qv)$ by Eq.\
(\ref{eq:D-axialG}) specialized to two dimensions:
\begin{subequations}
\label{G0s}
\begin{eqnarray}
G^0_{xx}(\qv) &=& {1 \over \lambda'_x  q_x^2 + K_y q_y^4 },
\label{G0x}\\
G^0_{yy}(\qv) &=& {1 \over \lambda'_y q_y^2 + K_x q_x^4 } ,
\label{G0y} \\
G^0_{xy}(\qv) &=& - {\lambda_{xy} q_x q_y\over 
\Delta_{\lambda} q_x^2 q_y^2 + K({\hat q})q^4 
( \lambda_y q_y^2 + \lambda_x q_x^2)} ,\ \ \ \ \ \ \ \ \ \ \ 
\label{G0xy1}
\end{eqnarray}
\end{subequations}
where
\begin{eqnarray}
\lambda'_{x,y}& = &\lambda_{x,y}\left(1 - {\lambda_{xy}^2\over
\lambda_x \lambda_y}\right) .
\end{eqnarray}
Stability requires $\lambda_x>0$, $\lambda_y >0$, and $\lambda_x
\lambda_y - \lambda_{xy}^2 >0$. $\lambda_x'$ and $\lambda_y'$ both
go to zero linearly in $\lambda_x \lambda_y - \lambda_{xy}^2$. The
dominant parts of the $u_x$ and $u_y$ correlation functions are
identical in form to the displacement correlation functions of a
two-dimensional smectic, which have been extensively studied\cite{salditt}.
Both $\langle u_x^2 \rangle$ and $\langle u_y^2 \rangle$ diverge
with system size:
\begin{eqnarray}
\langle u_x^2 \rangle &=& 4\int_{2 \pi/L_x}^{\infty}\int_{2 \pi
/L_y}^\infty {d
q_x d q_y\over (2 \pi )^2} G_{xx}^0 ( \qv ) \nonumber \\
& = & {1 \over \lambda'_x } \sqrt{L_x \over 2 \pi
a_x} \psi_u( 2 \pi a_x L_x /L_y^2 ) \\
& = & \begin{cases}
{1 \over \sqrt{2} \pi\lambda'_x }\sqrt{L_x \over
2 \pi a_x}, & \text{if $L_y^2 \gg 2 \pi a_x L_x$; } \\
{1 \over (2
\pi)^2} {1 \over \lambda'_x }{L_y \over a_x}, &\text{if $L_y^2 \ll
2 \pi a_x L_x$ ,}
\end{cases}
\nonumber
\end{eqnarray}
where $\psi_u(z)$ is a crossover function. The expression for
$\langle u_y^2 \rangle$ is obtained by interchanging $x$ and $y$
in the equation for $\langle u_x^2 \rangle$. The anisotropy
lengths $a_x$ and $a_y$ defined as:
\begin{subequations}
\label{anilength}
\begin{eqnarray}
a_x=(K_x/\lambda'_x)^{\frac{1}{2}}, \label{anilengthx}\\
a_y=(K_y/\lambda'_y)^{\frac{1}{2}}. \label{anilengthy}
\end{eqnarray}
\end{subequations}
They diverge as the stability limit $\lambda_x \lambda_y =
\lambda_{xy}^2$ is approached.

The connected phonon correlation functions at two spatially
separated points is
\begin{equation}
\label{Cij0}
C^0_{ij}(\rv)= \langle(u_i(\rv) -
u_i(0))(u_j(\rv)-u_j(0))\rangle_0
\end{equation}
For an infinite membrane
\begin{eqnarray}
C^0_{xx}(\rv)&=&2\int \frac{dq_x dq_y}{(2 \pi)^2}
         G^0_{xx}(\qv) (1 - e^{i \qv \cdot \rv}) \label{C0xxa}\nonumber\\
&=& {1 \over \lambda'_x }\sqrt{|x|\over \pi a_x} e^{- y^2/(4
a_x |x|)} \nonumber \\
& & + {1 \over 2 \lambda'_x }{|y| \over a_x} {\rm
erf}\left({|y|\over 2 \sqrt{a_x
|x|}}\right) , \nonumber\\
&\sim&\begin{cases}
{1 \over \lambda'_x }\sqrt{|x|\over \pi a_x} & \text{if
$|y|\ll 2 \sqrt{a_x |x|}$,} \\
{1 \over \lambda'_x } {|y|\over a_x} & \text{if
$|y|\gg 2 \sqrt{a_x |x|}$.} \\
\end{cases}
\label{C0xxb}
\end{eqnarray}
where ${\rm erf}(x)$ is the error function. The strong power-law
growth of $G^0_{ii}(\rv)$ indicates that thermal fluctuations lead
to arbitrarily large relative displacements of two points distant
on the membrane, again contrasting with the usual logarithmic
growth with $r$ in two-dimensional ordered $xy$-like systems.

The stability of the spontaneous nematic (uniaxial) order can be
analyzed by examining the rms fluctuations of the nematic director
field $\delta{\mathbf n}(\rv)$. Since in the nematic state, the
director is ``massively'' tied to the {\em antisymmetric} part of
the displacement gradient tensor $\eta_{ij}=\ppj
u_i$\cite{elast_us}, orientational fluctuations can be computed
from those of phonon fluctuations.  To linear order in rotations
of the director $\nv$ away from its preferred orientation along
the $x$-axis, $\delta n_y = \theta = \case{1}{2}(\partial_x u_y -
\partial_y u_x)$.  The angle correlation function is then
\begin{equation}
G_{\theta \theta} ( \qv ) = \frac{1}{4}\left[q_y^2 G^0_{xx}(\qv)+
q_x^2 G^0_{xx}(\qv)
  -2 q_x q_y G^0_{xy}(\qv)\right].
\label{dn02}
\end{equation}
It is clear from Eqs.\ (\ref{G0x})-(\ref{G0xy1}) that within the
harmonic approximation in-plane orientational fluctuations are
finite.

Turning to membrane out-of-plane fluctuations, we find that at
harmonic level local rms undulations $\langle \hv(\rv)^2 \rangle_0$
behave the same as those of polymerized membranes with
\begin{eqnarray}
\langle h(\rv)^2 \rangle_0
  &=& d_c \int_{\frac{2 \pi}{L_x}}\int_{\frac{2 \pi}{L_y}}
    \frac{dq_x dq_y}{2 \pi}  G^0_{h}(\qv)\nonumber \\
 & = &\frac{d_c (2 \pi L_x)^2}{\kappa} \psi_h(\frac{L_x}{L_y}),
\end{eqnarray}
where for simplicity we specialized the case of isotropic bending
rigidity $\kappa$ and defined scaling function
\begin{equation}
\psi_h(z) = 4 \int_1^{\infty} dx \int_z^{\infty} dy
  \frac{1}{(x^2+y^2)^2},
\end{equation}
with crossover property
\begin{equation}
\psi_h(z) \longrightarrow \left\{
             \begin{array}{ll} 2 \pi  & z \rightarrow 0\\
                            \frac{2 \pi}{z^2} & z \rightarrow \infty.
                               \end{array}\right.
\end{equation}

As in crystalline membranes, the strong $L^2$ growth implies
instability of the flat phase to thermal fluctuations, as well as the
importance of anharmonic elasticities, which (as in polymerized
membranes) can stabilize the flat phase.  Furthermore, in contrast to
polymerized membranes, here the power-law divergent smectic-like
in-plane phonon correlations that we found above, Eq.\ (\ref{C0xxb}),
suggest that phonon elastic nonlinearities in Eq.\ (\ref{uij}) may be
important as well. We turn to these questions in the next two
sections.

\section{Anharmonicities and the $\epsilon$-Expansion}
\label{AnharmonicModel}

As discussed in Sec.\ \ref{model}, rotational invariance in the target
space requires the elastic free energy to be expressed in terms of
nonlinear rather than linear strains.  The result is that there are
anharmonic couplings in the elastic energy both among phonon fields
and between phonon and height fields.  The violent power-law
fluctuations of these fields, the height field in particular, leads as
in other membranes systems\cite{KKN}-\cite{LR} to divergences with
system size $L$ of perturbations in anharmonic couplings for membranes
with spatial dimension $D$ less than a critical value $D_c$.  In this
section, we consider the interplay between fluctuations and anharmonic
couplings in $D$-dimensional nematic elastomer membranes embedded in a
$d$-dimensional space and show that perturbation theory breaks down
below $D_c=4$. We then study the anomalous elasticity of these
membranes in an $\epsilon$-expansion about $D=4$.  Our interest is in
developing insight into properties of real two-dimensional membranes.
We, therefore, consider only models that have a straightforward
analytic continuation to $D=2$.  The two models we consider are the
fully anisotropic $D$-axial elastomer, which has no surviving shear
modulus, and the uniaxial model in which we set the shear modulus
$\mu_{\perp}$ Eq.\ (\ref{nem-energy}) for shears in the plane perpendicular to
the nematic direction to zero.  These two models are equivalent in the
physical limit of $D=2$.

\subsection{Perturbative analysis of elastic nonlinearities}
\begin{figure}[!hbp]
\begin{center}
\includegraphics[width=8cm]{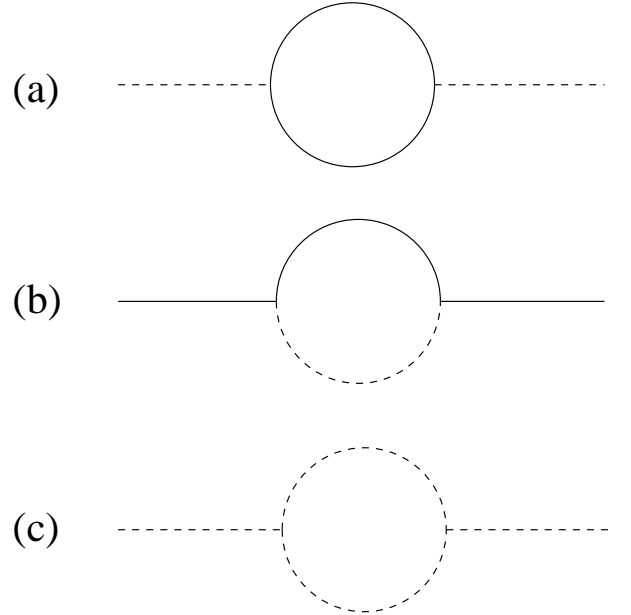}
\vspace{0.5cm}
\caption{Feynman diagrams renormalizing elastic
   constants $\lambda$'s (A), $\kappa$'s (B) and
  $K$'s (C). Solid lines denote undulation fields $\hv$,
  dashed lines denote phonon fields $\uv$.}
\label{Feynmandiagrams}
\end{center}
\end{figure}

The elastic free energy $\calF$, Eq.\ (\ref{Ftot}) contains
nonlinearities associated with membrane undulations (involving $\hv$
field) and in-plane phonon nonlinearities. The importance of
undulation nonlinearities is a consequence of membrane's vanishing
tension (softness of out-of-plane undulations, controlled by
curvature, rather than surface tension energy). As in an isotropic
polymerized membrane\cite{NP,AGL,LR}, undulation nonlinearities become
relevant when the dimension of the reference space is lower than 4.
To illustrate this point, we calculate the perturbative corrections to
elastic constants $\lambda_{ij}$ from undulation nonlinearities which
are represented by diagram (a) shown in Fig.  \ref{Feynmandiagrams},
\begin{eqnarray}
\delta \lambda_{ij}^h &=&
                - \frac{1}{2}\sum_{k,l}^D \lambda_{ik} \lambda_{jl}
                    \int \frac{d^Dq}{(2 \pi)^D}q_k^2 q_l^2(G^0_{h}(\qv))^2
                           \nonumber\\
   &=&      - \frac{d_c L^{4-D}}{2 (4-D)}
                  \sum_{k,l}^D \lambda_{ik} \lambda_{jl}
                    \int \frac{d\Omega_D}{(2 \pi)^D}\label{d_lambda}
\frac{\hat{q}_k^2 \hat{q}_l^2}{\kappa^2 ({\hat q})},\ \ \ \ \ \ \ \ \ \
\end{eqnarray}
where we assumed that the system has an equilibrium configuration
of $D$ dimensional sphere (disk in two dimensional case) with
radius $L$.  $d \Omega_D$ is the differential surface element of
the $D$ dimensional unit sphere. Quite clearly, for $D<4$, the
fluctuation corrections $\delta \lambda_{ij}$ diverge with system
size $L$. The associated nonlinear length scale, beyond which these
corrections become comparable to $\lambda_{ij}$, and perturbation
method breaks down is given by
\begin{equation}
L^h_{NL}\simeq \left( \frac{(4-D) \bar{\kappa}^2 }{ \lambda}
\right) ^{\frac{1}{4-D}},
\end{equation}
where $\lambda$ is the typical value of $\lambda_{ij}$ and
$1/\bar{\kappa}^2$ is defined as the angular integral 
in Eq.\,(\ref{d_lambda}).


If the modulus $\mu_{\perp}$ for shears in the direction perpendicular to
the anisotropy axis in the uniaxial model is zero, then
fluctuations in $\uv_t$ Eq.\ (\ref{u-rep}) have the same
$q^{-4}$ harmonic-theory divergence as height fluctuations, but
with $K({\hat q})$ replacing $\kappa({\hat q})$.  Thus, when
$\mu_{\perp}=0$, both $\vec h$ and $\uv_t$ contribute to divergences in
$\lambda_{ij}$ below $D=4$.  This point was missed in the analysis
of the fixed connectivity fluid fixed point in Ref.\
\onlinecite{AGL}.

As discussed in the Introduction, the spontaneously broken
in-plane rotational symmetry of the nematic elastomer membrane
leads to soft in-plane elasticity.  As a result, in strong
contrast to isotropic or crystalline membranes, in-plane
nonlinearities in $\calF$, Eq.\ (\ref{Ftot}), also correct elastic
moduli $\lambda_{ij}$. Its contribution $\delta \lambda_{ij}^u$,
as represented by diagram (c) in Fig.\ \ref{Feynmandiagrams}, is
given by
\begin{equation}
\delta \lambda_{ij}^u = - \frac{1}{2}\sum_{k,l}^D
     \lambda_{ik}\lambda_{jl}\int \frac{d^Dq}{(2 \pi)^D} q_k^2 q_l^2
         \sum_{mn}^D |G^0_{mn}(\qv)|^2.
\end{equation}
In the $D$-axial model, the dominant fluctuations in $G_{nm}^0$ are
smectic-like, and it is straightforward to show that they cause the
above correction to $\lambda_{ij}^u$ to diverge as $L^{(3-D)/2}$ below
$D=3$ if the phonon fluctuations retained their harmonic character
down to $D=3$. In contrast in the uniaxial (analytical continuation)
model, the contributions to $\delta \lambda_{ij}^u$ due to $\uv_t$
fluctuations diverge below $D=4$, when $\mu_{\perp}=0$ as discussed above.
The $nn$ part of $G_{nm}^0$ is, however, smectic-like and diverges
more weakly as $L^{(3-D)/2}$.

Similar calculation shows that the perturbative correction to in-plane
elastic constants $K_{ij}$ is dominated by in-plane nonlinearities and
also diverges below three dimensions.  Thus the upper critical
dimension for undulation nonlinearities is 4, while in-plane
nonlinearities become relevant below $D=3$.  For $3<D<4$, in-plane
nonlinearities are irrelevant, consequently in-plane curvature energy
moduli $K_{ij}$ are only renormalized finitely.

Undulation nonlinearities renormalize bending rigidities
$\kappa_{ij}$ as well as in-plane elastic constants
$\lambda_{ij}$. One would expect that as in the case of
isotropic polymerized membranes, the perturbative corrections
to $\kappa_{ij}$ also diverge as $L^{4-D}$. This is,
however, not true, as one can see from the following
argument. Note that if we neglect in-plane nonlinear
terms (which is legitimate above 3 dimension) in 
Eq.\ (\ref{Ftot}), and set $K_{ij}$ to be zero, we can integrate out the
phonon fields completely and what is left is only the bending
energy $\sum_{ij}\kappa_{ij} (\ppi^2 \hv \cdot \ppj^2 \hv)$
without any nonlinearities.  Thus there should be no
anomalous elasticity for bending rigidity $\kappa_{ij}$, were
there no in-plane curvature rigidities $K_{ij}$. This implies
that, with the presence of $K_{ij}$, the renormalization of
bending rigidities $\kappa_{ij}$ is dominated by anisotropic,
smectic-like modes, for which in-plane curvature rigidities
$K_{ij}$ are important, and the critical dimension below which
$\kappa_{ij}$ is infinitely renormalized is 3, as in smectic
liquid crystals.



The upshot of above discussion is that near $D=4$ in the $D$-axial
model all in-plane nonlinearities in $\uv$ are irrelevant, and in the
uniaxial model nonlinearities in $u_n$ and $u_L$ are irrelevant.
Consequently, to capture the long-wavelength behavior we can use
simplified expressions for the full non-linear strain tensors given by
\begin{subequations}
\label{reduceduij}
\begin{eqnarray}
\hspace{-5mm}
\lefteqn{D-{\rm axial}:}\nonumber \\
u_{ij} &\rightarrow &\case{1}{2}(\partial_i u_j +
\partial_j u_i +
\partial_i \hv \cdot \partial_j \hv ),\\
\lefteqn{\rm uniaxial:} \nonumber \\
u_{ij} &\rightarrow
&\case{1}{2}(\partial_i u_j +
\partial_j u_i +
\partial_i \hv \cdot \partial_j \hv +\partial_i \uv_t
\cdot \partial_j \uv_t ).
\end{eqnarray}
\end{subequations}
The effective Hamiltonian for studying membranes without a shear
modulus is the $\calF$ Eq.\ (\ref{Ftot}) with $\mu_{\perp} =0$ and one
of the above reduced strains.

The difference between this model free energy and that of isotropic
polymerized membranes\cite{NP,AGL}, as well as that of
fixed-connectivity fluid membranes\cite{AGL}, should be stressed.
First, the bare elastic constants, $\kappa_{ij}$ and $\lambda_{ij}$
are anisotropic rather than isotropic, and the anisotropy in
$\lambda_{ij}$ cannot be eliminated by a simple rescaling of lengths.
Second and more importantly, the energy cost for shear in the planes
of anisotropy is {\em zero} because of the spontaneous broken symmetry
of the nematic state. Thus, our model free energy, Eq.\ (\ref{Ftot}),
is not a simple anisotropically scaled generalization of
fixed-connectivity fluid\cite{AGL}. The matrix of coupling constants
$\lambda_{ij}$ is such the elastic energy can not be reduced to that
of density variation alone.

\subsection{Renormalization Group and $4-D$-expansion}
\label{RG}

In isotropic membranes below $D=4$, height fluctuations lead to
anomalous elasticity\cite{KKN}-\cite{LR} with bending modulus and
elastic modulus, respectively, diverging and vanishing with
wavenumber as
\begin{equation}
\kappa ( \qv ) \sim q^{-\eta_\kappa} , \qquad \lambda( \qv ) \sim
q^{\eta_{\lambda}} ,
\end{equation}
where the exponents $\eta_{\kappa}$ and $\eta_{\lambda}$ are
related via the Ward identity\cite{AGL}
\begin{equation}
2 \eta_{\kappa} + \eta_{\lambda} = \epsilon = 4 - D .
\end{equation}
We have just argued that $\kappa$ is not renormalized for dimensions
near 4 in the models without shear moduli that we are considering.
Thus $\eta_{\kappa}=0$, and $\eta_{\lambda} = \epsilon$.  In this
section, we will show explicitly within an RG calculation that this is
indeed the case in both models we consider.

We use standard momentum-shell renormalization-group procedures,
integrating out a shell in momentum space with $\Lambda/e^l < q <
\Lambda$, where $\Lambda$ is the ultraviolet cutoff, to produce
fields $\uv^< ( \rv )$ and $\hv^< ( \rv )$ with wave-numbers with
magnitude $q < e^{-l} \Lambda$.  We then rescale lengths and
fields according to
\begin{eqnarray}
\rv&=&\rv'e^l,\\
u_i^< (\rv)&=&e^{\chi_i l} u_i'(\rv'),\\
\hv^<(\rv)&=&e^{\phi l}\hv'(\rv'),
\end{eqnarray}
so as to restore the ultraviolet cutoff to its original value
$\Lambda$. In the uniaxial case, we decompose $\uv$ as in Eq.\ 
(\ref{u-rep}) and choose $\chi_n = \chi_L= \chi$ and $\chi_t$
different from $\chi$.  In the $D$-axial case, we can chose all of the
$\chi_i$ to be equal to $\chi$. It is convenient to choose the
rescaling so as to preserve the nonlinear form of the strain $u_{ij}$
Eq.\ (\ref{reduceduij}). This requires
\begin{equation}
\chi = 2 \phi - 1 = 2 \chi_t -1
\end{equation}

The integration over the high wave-vector components of $\uv$ and
$\hv$ can be performed perturbatively in nonlinearities of
Hamiltonian, in a procedure very similar to that of perturbative
analysis.  The Feynman graph giving the 1-loop corrections to
$\lambda_{ij}$ is shown in Fig.\ \ref{Feynmandiagrams}, diagram
(a).

After performing the rescaling and calculating the graphic
corrections, we obtain the following RG flow equations:
\begin{subequations}
\begin{eqnarray}
\frac{d\lambda_{ij}}{dl}
      &=& (D-2+2 \chi)\lambda_{ij} -
               \case{1}{2} \sum_{k,l} \lambda_{ik} M_{kl}
    \lambda_{lj},\ \ \ \ \ \ \ \ \ \ \ \ \   \label{flowlambda}\\
\frac{d\kappa_{ij}}{dl} &=& (D-4+2 \phi) \kappa_{ij}, \\
\frac{d K_{ij}^n }{dl} & = & ( D - 4 + 2 \chi) K_{ij}^n,\\
\frac{d K_{ij}^t }{dl} & = & ( D - 4 + 2 \chi_t) K_{ij}^t ,
\label{eq:recur1}
\end{eqnarray}
\end{subequations}
where, for simplicity, we have set the ultra-violet cutoff $\Lambda
=1$.  The components of the in-plane bending coefficient, $K_{ij}^n$,
coupling to $u_n$ and $u_L$ scale with $\chi$, whereas those,
$K_{ij}^t$, coupling to $\uv_t$ scale with $\chi_t$. The matrix
$M_{kl}$ has similar but different forms for the uniaxial and
$D$-axial models. In the $D$-axial model, $M_{kl} = d_c
M_{kl}^{\kappa}$, and in the uniaxial model $M_{kl} = d_c
M_{kl}^{\kappa} + (D-2) M_{kl}^K$, where
\begin{subequations}
\label{eq:MK}
\begin{eqnarray}
M_{kl}^{\kappa} &= & \int {d \Omega_D \over (2 \pi)^D} {{\hat
q}_k^2 {\hat q}_l^2 \over \kappa^2 ( {\hat q})} \\
M_{kl}^K & = & \int {d \Omega_D \over (2 \pi)^D} {{\hat
q}_k^2 {\hat q}_l^2 \over K^2( {\hat q})} .
\end{eqnarray}
\end{subequations}
Note that $M_{kl}$, like $\lambda_{ij}$ and $A_{ij}$ defined
in Sec. \ref{model}, is really a subset of the components
constructed from the fourth-rank tensor
\begin{equation}
M_{ijkl}^{\kappa} = \int {d \Omega_D \over (2 \pi)^D}{{\hat
q}_i {\hat q}_j {\hat q}_k {\hat q_l}\over \kappa^2({\hat q})
q^4}
\label{M-4}
\end{equation}
and $M^K_{ijkl}$ defined in a similar way. $M_{kl}$ does not
transform like nor have the symmetries of a second-rank tensor.
The $D$-axial and the uniaxial model differ mostly in their
respective forms of $\lambda_{ij}$ and $M_{ij}$. These different
forms require slightly different fixed-point analysis, and we will
now treat the two cases separately.

Regardless of the model, we can choose $\phi = (4-D)/2$ and $\chi_t =
(4-D)/2$ to keep $\kappa_{ij}$ and $K_{ij}^t$ fixed.  Then, $\chi =
3-D$, and the inverse correlation function for the parts of $\uv$ not
in the anisotropy planes (i.e., not $\uv_t$) scales as
\begin{eqnarray}
G^{-1}_{ij} ( \qv , \lambda_{ij} , K_{ij} ) &=&
e^{(D-6)l}G_{ij}^{-1}(e^l \qv, \lambda_{ij} (l), K_{ij} ( l ) )
\nonumber\\
& = & e^{-\epsilon l}\lambda_{ij} (l) q_i q_j + K ({\hat q}) q^4 .
\end{eqnarray}
Thus, if $\lambda_{ij}$ has a non-zero fixed point value
$\lambda_{ij}^*$, then, choosing $e^{l} q = 1$, we have
\begin{equation}
\lambda_{ij}(q) = \lambda_{ij}^* q^{\epsilon}.
\end{equation}
Both $\kappa_{ij}$ and $K_{ij}^t$ remain constant.

\subsection{The Uniaxial Model}

In the uniaxial model, both $\lambda_{ij}$ Eq.\
(\ref{lambda-uni}) and $M_{ij}$ are uniaxial.  $M_{ij}$ is easily
calculated by taking the appropriate components of the full
fourth-rank tensor $M_{ijkl}$ Eq.\ (\ref{M-4})
\begin{eqnarray}
M_{ij} &= & M_{||} \delta_{i1} \delta_{j1}  \\
& & + {1 \over (D-1)} M_{||\perp}[\delta_{i1} (1 - \delta_{j1}) +
(1- \delta_{i1})\delta_{j1}] \nonumber\\
& & + {1 \over (D-1)(D+1)}M_{\perp}(1+ 2 \delta_{ij})(1-
\delta_{i1})(1- \delta_{j1}) .\nonumber
\end{eqnarray}
The components of $M_{ij}^{\kappa}$ and $M_{ij}^K$ of $M_{ij}$
have simple expressions in terms of integrals over angle:
\begin{subequations}
\begin{eqnarray}
M_{||}^{\kappa} & = & \int {d \Omega_D\over (2 \pi )^D}{\cos^4
\theta \over \kappa^2({\hat q})} \\
M_{||\perp}^{\kappa}& = &\int {d \Omega_D\over (2 \pi )^D}{\cos^2
\theta
\sin^2 \theta \over \kappa^2({\hat q})} \\
M_{\perp}^{\kappa} & = & \int {d \Omega_D\over (2 \pi )^D}{\sin^4
\theta \over \kappa^2({\hat q})} ,
\end{eqnarray}
\end{subequations}
where $\theta$ is the angle ${\hat q}$ makes with the uniaxial axis
($1$-axis).  A similar set of expressions applies to the
components of $M_{ij}^K$ with $K({\hat q})$ replacing $\kappa(
{\hat q})$. Note that $M_{23} = M_{\perp}/[(D-1)(D+1)]$ and
 $M_{12} = M_{13} = M_{|| \perp}/(D-1)$, whereas the
components of the true second rank uniaxial tensor $T_{ij}$
satisfies $T_{23} = T_{12} = T_{13}=0$.

\begin{figure}
\includegraphics{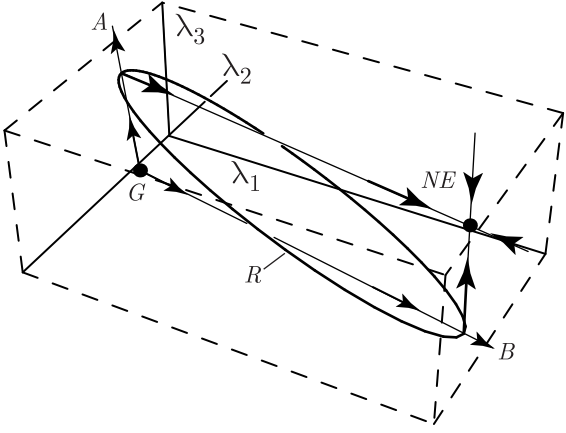}
\caption{Renormalization group flows for the uniaxial model,
  showing the Gaussian ($G$) and uniaxial nematic elastomer ($NE$)
  fixed points and the fixed ring $R$, unstable in one direction and
  marginally stable around its perimeter. The locus of flow lines, of
  which $GA$ and $GB$ are two examples, from $G$ to $R$ and extended
  beyond $R$ form a distorted cone $C$.  All points within and on the
  boundary of $C$ flow to the stable fixed point $NE$.  All other
  points flow to large coupling.}
\label{fig:uniaxial}
\end{figure}

With these definitions, the recursion relations for the components
of $\lambda_{ij}$ become
\begin{subequations}
\label{eq:lambda-recur1}
\begin{eqnarray}
\hspace{-10mm}
\frac{d \lambda_1}{dl} & = & \epsilon \lambda_1 - \case{1}{2}
(\lambda_1^2 M_{||} + 2 \lambda_1 \lambda_2
M_{||\perp} + \lambda_2^2 M_{\perp}), \\
\frac{d \lambda_2}{dl}&= &\epsilon \lambda_2 -
\case{1}{2}(\lambda_1 \lambda_2 M_{||} + \lambda_2^2
M_{||\perp}\nonumber \\
& & +\lambda_1 \lambda_3
M_{||\perp} + \lambda_2 \lambda_3 M_{\perp}),\\
\frac{d \lambda_3}{dl} &= &\epsilon \lambda_3 -
\case{1}{2}(\lambda_3^2 M_{\perp} + 2 \lambda_2\lambda_3
M_{||\perp} + \lambda_2^2 M_{||}),
\end{eqnarray}
\end{subequations}
where $\lambda_1$, $\lambda_2$, and $\lambda_3$ were
defined in Eq.~(\ref{lambda-uni}).
These equations have an unusual fixed-point structure as
shown in Fig.\ \ref{fig:uniaxial}. In appropriate
two-dimensional planes in the three-dimensional space of
$\lambda_1$, $\lambda_2$, and $\lambda_3$, they exhibit the
familiar four-fixed-point structure of systems with two
coupled potentials in which there is one unstable Gaussian
fixed point, one globally stable fixed point, and two fixed
points that are stable in one direction and unstable in the
other.  The full three-dimensional structure is topologically
equivalent to what would be obtained if the two-dimensional
structure is rotated about the axis connecting the Gaussian
and globally stable fixed point and subsequently stretched
anisotropically. In this process, the two mixed-stability
fixed points become an elliptical ring with stability
exponent of zero for displacements along the ring and one
positive and one negative exponent for displacements
perpendicular to the ring.  The fixed points and their
stability exponents $\omega_1$, $\omega_2$, and $\omega_3$
are
\begin{enumerate}
\item Gaussian
\begin{eqnarray}
\lambda_1& = &\lambda_2 = \lambda_3=0, \\
\omega_1 &= &\omega_2 = \omega_3 = \epsilon. \nonumber
\end{eqnarray}
\item Uniaxial nematic elastomer
\begin{eqnarray}
\hspace{-5mm} \lambda_1 & = & { 2 \epsilon M_{\perp}\over
\Delta_M}, \qquad \lambda_2 =  -{2 \epsilon M_{||\perp}\over
\Delta_M}, \qquad
\lambda_3 = {2 \epsilon M_{||}\over \Delta_M} , \nonumber\\
\omega_1& =& - \epsilon, \qquad \omega_2 = -\epsilon , \qquad
\omega_3 = -\epsilon \label{fp_uniaxial} ,
\end{eqnarray}
where $\Delta_M = M_{||} M_{\perp} - M_{||\perp}^2$.
\item Fixed Ring
\begin{eqnarray}
\lambda_1 & = & {2 \epsilon \over M_{||} + 2 \alpha M_{||\perp} +
\alpha^2 M_{\perp}}, \nonumber \\
\lambda_2 &= &\alpha \lambda_1, \qquad
\lambda_3 = \alpha^2 \lambda_1 \\
\omega_1 & = & - \epsilon,\qquad \omega_2 = 0 , \qquad \omega_3 =
 \epsilon ,\nonumber
\end{eqnarray}
for $-\infty \leq \alpha < \infty$.  For every point on the ring,
$\lambda_1 \lambda_3 = \lambda_2^2$.  The fixed ring includes
various interesting points:
\begin{enumerate}
\item $\alpha=1$, $\lambda_1 = \lambda_2 = \lambda_3$.  This is the
  fully isotropic fixed-connectivity-fluid fixed point of isotropic
  membranes\cite{AGL}.
\item $\alpha = -1$, $\lambda_1= - \lambda_2 = \lambda_3$.
\item $\alpha = \pm \infty$, $\lambda_1 = \lambda_2=0$, $\lambda_3
= 2 \epsilon/M_{\perp}$.
\item $\alpha=0$, $\lambda_1=2 \epsilon/M_{||}$, $\lambda_2 =
\lambda_3 = 0$.
\end{enumerate}
\end{enumerate}

\subsection{The $D$-Axial Model}

The analysis of the $D$-axial model is complicated by the
fact that $\lambda_{ij}$ has a large number ($D(D+2)/2 = 10$
in $D=4$) of independent components. To simplify our
presentation, we will first consider flows in a restricted
subspace in which $\lambda_{ij}$ is parameterized by only two
parameters. Thus we will first find 4 fixed points with
particular structure and show that one of them is globally
stable and describes $D$-axial nematic elastomer membrane. We
then give a general solution for all fixed points and show
that it actually contains both the fixed point Eq.\
(\ref{fp_uniaxial}) and the fixed ring we found for uniaxial
model. Furthermore, we show that if anisotropy of
$\lambda_{ij}$ is turned on in the $(D-1)$-dimensional plane,
the uniaxial fixed point
 Eq. (\ref{fp_uniaxial}) become unstable and the system flows to
the globally stable D-axial one.

Our recursion relations are still given by Eq.\ (\ref{eq:recur1})
without $K_{ij}$.  Choosing $\phi$ to keep $\kappa_{ij}$ constant,
we obtain
\begin{equation}
{d \mm{\lambda}\over dl} = \epsilon \mm{\lambda} -
\case{1}{2}\mm{\lambda} \,\mm{M}\, \mm{\lambda} ,
\label{lambda-recur}
\end{equation}
where $\mm{\lambda}$ and $\mm{M}$ are, respectively, matrices with
entries $\lambda_{ij}$ and $M_{ij}$.  As in the uniaxial case,
this equation should have an isotropic fixed point will all
$\lambda_{ij}$ equal, i.e., with $\mm{\lambda} \sim \mm{1}$ where
$\mm{1}$ is the matrix with all entries equal to one. It is also
clear that this equation has fixed point with $\mm{\lambda} \sim
\mm{M}^{-1}$ provided the symmetries of $\mm{\lambda}$ and
$\mm{M}$ are compatible. In the uniaxial case, $\mm{M}$ has more
unequal components than $\mm{\lambda}$, thus
the solution $\mm{\lambda} \sim \mm{M}^{-1}$ is not permitted there.
In the $D$-axial case, $\mm{M}$ has the same number of independent
components as a $D$-axial second-rank tensor, and
$\mm{\lambda}\sim \mm{M}^{-1}$ is permitted.  We thus begin by
seeking fixed points in the $2D$ subspace defined by
\begin{equation}
\mm{\lambda} = \lambda_a \mm{M}^{-1} + \lambda_b \mm{1} .
\label{lambdares}
\end{equation}
The recursion relations for $\lambda_a$ and $\lambda_b$ are
\begin{subequations}
\begin{eqnarray}
{d \lambda_a \over dl} & = & \epsilon \lambda_a - \case{1}{2}
\lambda_a^2 \\
{d \lambda_b\over dl} & = & \epsilon \lambda_b - \case{1}{2}
\lambda_b ( 2 \lambda_a + \lambda_b \sum_{i,j=1,}^D M_{ij} ) ,
\end{eqnarray}
\label{eq:lambda-recur2}
\end{subequations}
where we used $\mm{1}\, \mm{M}\,\mm{1} = \sum_{ij}^D M_{ij}
\mm{1}$. The fixed points and their stability exponents for these
equations are
\begin{enumerate}
\item Gaussian
\begin{eqnarray}
\lambda_a &= &\lambda_b = 0 \\
\omega_a & = &\omega_b = \epsilon .
\end{eqnarray}
\item Fixed-connectivity fluid
\begin{eqnarray}
\lambda_a & = &0, \qquad \lambda_b = {2 \epsilon \over \sum_{ij}^D
M_{ij}} \\
\omega_a &= &\epsilon, \qquad \omega_b = - \epsilon .
\end{eqnarray}
This fixed point is in fact the isotropic fixed-connectivity
fluid fixed point found in Ref.\ \cite{AGL} for isotropic
membranes.
\item Mixed
\begin{eqnarray}
\lambda_a &= &2 \epsilon, \qquad \lambda_b = - {2 \epsilon\over
\sum_{ij}^D M_{ij}} \\
\omega_a &= &- \epsilon , \qquad \omega_b = \epsilon .
\end{eqnarray}
We call this fixed point mixed because its coupling constant
matrix $\mm{\lambda}$ has components characteristic of both
the fixed connectivity fluid and the $D$-axial membrane fixed
points.
\item $D$-axial nematic elastomer
\begin{eqnarray}
\lambda_a &= &2 \epsilon, \qquad \lambda_b = 0 , \\
\omega_a & = & \omega_b = - \epsilon .
\end{eqnarray}
This fixed point is in fact globally stable, i.e., it is stable in
all directions. Plugging $\mm{\lambda} = \lambda_a \mm{M}^{-1}$
into the recursion relation Eq.\ (\ref{lambda-recur}) and
linearizing, we obtain $d\delta\mm{\lambda}/dl = - \epsilon \delta
\mm{\lambda}$. Thus, the stability exponent is $-\epsilon$ for all
directions.
\end{enumerate}

We have thus identified one globally stable fixed point, the
nematic elastomer fixed point, and two others. The full fixed
point structure of the flow equation Eq.\ (\ref{lambda-recur})
is actually not difficult to determine. Since  $\mm{M}$ is
generically positive definite, we can define a new
(symmetric) coupling constant matrix $\mm{P}=\mm{M}^{\frac{1}{2}}
\mm{\lambda}\, \mm{M}^{\frac{1}{2}}/(2 \epsilon)$, whose flow
equation given by:
\begin{equation}
\frac{d}{dl} \mm{P} = \mm{P} - \mm{P}^2.\label{flow_P}
\end{equation}
One immediately see that every projection
matrix\footnote{A projection matrix is a symmetric matrix which equals
to the square of itself.} is a fixed point for
$\mm{P}$ and vice versa. This means that the general solution for
the flow equation Eq.\,(\ref{lambda-recur}) is given by:
\begin{equation}
\mm{\lambda} = 2 \epsilon \mm{M}^{-\frac{1}{2}} \mm{P} \, \mm{M}^{-\frac{1}{2}},
\end{equation}
with $\mm{P}$ an arbitrary projection matrix.

We can classify fixed points, or more generally fixed
subspaces, by the dimension $D_P$ of the space that $P$
projects onto.  Of course, ${\rm Tr} P = D_P$.  If the
dimension $D$ of the elastic manifold is an integer, then $P$
can project onto all subspaces with dimensions $D_P = 0,1,
... , D$.  When $D$ is not an integer, the classification is
less clear.  A convenient set, however, is the set with
dimensionalities $D_P = 0,1, ... , [D]$, and $D_P=D,D-1, ...
, D - [D]$, where $[D]$ is the greatest integer less than or
equal to $D$.  If $D_P = D$, $\mm{P}^D= \mm{I}$ is the unit
matrix, where the superscript indicates the dimension of $P$.
If $D_P = 1$, $P^1_{ij}=e_i e_j$ for {\em any} unit vector
$\ev$. A $(D-1)$-dimensional operator $P_{ij}^{(D-1)} =
\delta_{ij} -e_i e_j$ can also be constructed from the unit
vector $\ev$.  Similarly, $k$- and $(D-k)$-dimensional
projection operators can be defined via $P^k_{ij} =
\sum_{l=1}^k e_{li} e_{lj}$ and $P^{(D-k)}_{ij} = \delta_{ij}
- P^k_{ij}$, where $\ev_k\cdot \ev_l =\delta_{kl}$.



To study stability of the flow equation, Eq.\ (\ref{flow_P}), for $P$
for a given fixed-point projection matrix, $P_0^{D_0}$, we express
deviations of $P$ from $P_0^{D_0}$ as
\begin{equation}
\delta \mm{P} = \mm{P} - \mm{P}_0^{D_0}
 =  \mm{P}_1 + \mm{P}_2 + \mm{P}_3,
\end{equation}
where
\begin{eqnarray}
\mm{P}_1 &=& \mm{P}_0^{D_0} \delta \mm{P} \, \mm{P}_0^{D_0}, \\
\mm{P}_2 &=& (\mm{I} - \mm{P}_0^{D_0})\delta\mm{P}(\mm{I} - \mm{P}_0^{D_0}), 
\ \ \ \ \ \ \ \ \ \\
\mm{P}_3 &=& \mm{P}_0^{D_0} \delta \mm{P} (\mm{I} -
\mm{P}_0^{D_0})
 +  (\mm{I} - \mm{P}_0^{D_0}) \delta \mm{P} \, \mm{P}_0^{D_0}.\ \ \ \ \ \ \ \
\end{eqnarray}
Recall that $\mm{P}$ is a $D$-dimensional symmetric matrix
with $D(D+1)/2$ independent components.  $\mm{P}_1$ is the
projection of $\delta P$ onto the $D_0$-dimensional subspace
defined by $\mm{P}_0^{D_0}$, and it has $D_0(D_0+1)/2$
independent components.  Similarly, $\mm{P}_2$ is the
projection of $\delta \mm{P}$ onto the $D-D_0$ dimensional
subspace defined by $\mm{I}-\mm{P}_0^{D_0}$ with
$(D-D_0)(D-D_0+1)/2$ independent components. Finally,
$\mm{P}_3$ represents the $D_0(D-D_0)$ components of $\delta
\mm{P}$ that couple the subspaces defined by $\mm{P}_0^{D_0}$
and $\mm{I}-\mm{P}_0^{D_0}$. To linear order, the flow
equations for $\delta P$ are
\begin{eqnarray}
\frac{d}{d\,l} \mm{P}_1 &=& - \mm{P}_1,\\
\frac{d}{d\,l} \mm{P}_2 &=&  \mm{P}_2,\\
\frac{d}{d\,l} \mm{P}_3 &=& 0.
\end{eqnarray}
Thus, $\mm{P}_1$ is stable, $\mm{P}_2$ is unstable, and
$\mm{P}_3$ is marginally stable.  This means that the
$D_0$-dimensional fixed-point projection matrix is stable
with respect to any change in $\mm{P}$ in the $D_0 (D_0+1)/2$
dimensional space of $D_0 \times D_0$ matrices that operate
in the space defined by $\mm{P}_0^{D_0}$; it is unstable with
respect to any changes in $P$ in the $(D-D_0)
(D-D_0+1)/2$-dimensional space of $(D-D_0) \times (D-D_0)$
matrices that operate in defined by $\mm{I}-\mm{P}_0^{D_0}$;
and it is marginally stable with respect to changes in
$\mm{P}$ in the $D_0 (D-D_0)$-dimensional space of matrices
that couple $\mm{P}_0^{D_0}$ to $\mm{\delta}-\mm{P}_0^{D_0}$.
These results imply that the set of fixed points defined by
all $D_0$-dimensional projections matrices is a $D_0
(D-D_0)$-dimensional surface in the space of all possible
symmetric matrices $P$ or, equivalently in the space of
coupling constants $\mm{\lambda}$. This space is necessarily
compact since the subspaces defined by $\mm{P}_0^{D_0}$ are
parameterized by unit vectors.

There is only one projection operator with $D_0=D$.  This is
the operator $\mm{P}_0^D=\mm{I}$ that projects onto the whole
space.  For this case, $\mm{P_2}$ and $\mm{P_3}$ are both
zero, and the fixed point is stable in all directions. It is
the globally stable fixed point with $\mm{\lambda} = 2
\epsilon \mm{M}^{-1}$, which is identical to the stable
$D$-axial fixed point of the restricted set class of
couplings defined by Eq.\ (\ref{lambdares}).  The other fixed
points for the restricted set of couplings must correspond to
some $\mm{P}_0^{D_0}$ with $D_0<D$.  It is straightforward to
show that the fixed-connectivity-fluid fixed point
corresponds to ${P}_{0ij}^1= e_i e_j$ with
\begin{equation}
e_i = {\sum_{k=1}^D M_{ik}^{\frac{1}{2}}\over (\sum_{i,j=1}^D M_{ij}
)^{\frac{1}{2}}} ,
\end{equation}
where $M_{ij}^{\frac{1}{2}}$ is the $ij$ component of the matrix
$\mm{M}^{\frac{1}{2}}$.
Thus, this fixed point is actually a single point in a
$(D-1)=(3-\epsilon)$-dimensional fixed manifold. Similarly,
the mixed fixed point corresponds to ${P}_{0ij}^{D-1}=
\delta_{ij} - e_i e_j$.  There are other unstable fixed
points for $\mm{\lambda}$  not described by the restricted
set defined by Eq.\ (\ref{lambdares}), in particular those
with $D_0=2$ or $D-2$.

We have identified all of the fixed point manifolds of the
$D$-dimensional coupling matrix $\mm{\lambda}$.  These must
include the fixed points of the uniaxial model discussed in
the preceding subsection. When uniaxial constraints  are
applied, it is natural to construct unit vectors and
projection matrices from the components of $\mm{M}$ parallel
and perpendicular to the anisotropy axis.  It is not
difficult to show that the  stable uniaxial fixed point
corresponds to a two-dimensional projection operator
\begin{equation}
P_{0ij}^2 = e_{1i} e_{1j} + e_{2i} e_{2j} ,
\end{equation}
where
\begin{eqnarray}
e_{1i} & = & 
M_{i1}^{\frac{1}{2}}/M_{||}^{\frac{1}{2}}, \\
e_{2i} & = & \sqrt{M_{||}\over \Delta_M}\left(\sum_{k=2}^D
M_{ik}^{\frac{1}{2}}- {M_{||\perp}\over M_{||}} M_{1i}^{\frac{1}{2}}\right).
\end{eqnarray}
Thus, the uniaxial fixed point is a
point, satisfying uniaxial constraints, in a
$2(D-2)$-dimensional fixed manifold of all possible
couplings.  A point on the uniaxial fixed ring parameterized
by $\alpha$ corresponds to a one-dimensional projection
operator $\mm{P}_0^1$ defined by the unit vector
\begin{equation}
e_i ( \alpha ) = {M_{i1}^{\frac{1}{2}} + \alpha \sum_{k=2}^D
M_{ik}^{\frac{1}{2}} \over (M_{||} + 2 \alpha M_{||\perp} + \alpha^2
M_{\perp})^{\frac{1}{2}}} .
\end{equation}
The set of vectors $\ev({\alpha})$ defined by all $\alpha$
define a one-dimensional loop in a $(D-1)$-dimensional fixed
manifold in the space of all possible $D$-dimensional
couplings.

\section{Discussion and Conclusion }
\label{Conclusion}

\begin{figure}[!hbp]
\label{phasediagram}
\begin{center}
\includegraphics[width=7cm,height=12cm]{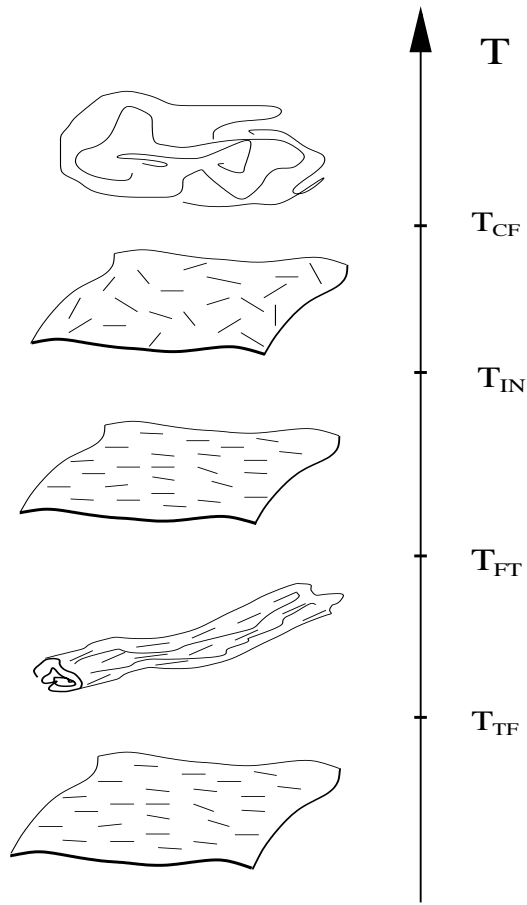}
\vspace{0.5cm}
\caption{A possible phase diagram for ideal nematic elastomer
membranes.  As the temperature is lowered a crumpled membrane
undergoes a transition to isotropic flat phase at $T_{CF}$, followed
by a 2D in-plane isotropic-nematic like transition to an
anisotropic(nematic) flat phase.  As $T$ is lowered further, this
anisotropic flat phase becomes unstable to a nematic tubule
 phase, where it continuously crumples
in one direction but remains extended in the other. At even lower temperature,
a tubule-flat transition takes place at $T_{TF}$.}
\label{tubule}
\end{center}
\end{figure}

In this paper we studied thermal fluctuation and nonlinear elasticity
of nematically-ordered elastomer membranes in their flat phase.  For
the physical case of two-dimensional membranes, we found that at
harmonic level in-plane phonon correlations are short-ranged but the
in-plane orientational order remains long-range, in spite of violent
thermal fluctuations. A generalization of a nematic elastomer membrane
to an arbitrary $D$-dimensional membrane with either uniaxial or
$D$-axial nematic order, allowed us to study the effects of
out-of-plane undulation and in-plane phonon nonlinearities. We found
that undulation nonlinearities are relevant $D<4$ and dominate over
the in-plane nonlinearities that only become important
$D<3$. Focusing on the dominant undulation nonlinearities (and
neglecting in-plane nonlinearities) we performed an RG calculation
combined with an expansion about $D=4$ and found that for $3<D<4$,
bending rigidities are only finitely renormalized, while in-plane
elastic moduli become singular functions of a wavevector (i.e.,
exhibit anomalous elasticity), vanishing with a universal power-law.
This power-law is controlled by an infrared stable fixed point whose
stability we analyzed in detail for the uniaxial and $D$-axial
analytic continuations, finding agreement in fixed point
structure. This analysis also allowed us to make contact and recover
some of the results previously obtained in the studies of
(crystalline) polymerized membranes. In particular we found that the
so-called connected fluid is realized as a fixed point of a
nematically-ordered elastomer membrane that is unstable to the
globally stable nematic-elastomer fixed point.

Despite of some of the success in our understanding of the behavior of
nematic elastomer membranes, there are obvious limitations of our
analysis, most notably in the application of our work to the physical
case of $D=2$ elastomer membranes.  This shortcoming primarily has to
do with the neglect of in-plane elastic nonlinearities, which, near
the Gaussian fixed point become relevant for $D<3$. While it is very
likely that the subdominance of these in-plane nonlinearities relative
to the undulation ones will persist some amount {\em below} $D=3$
\cite{phi4comment}, we expect that in the physical case of $D=2$ all
three nonlinearities need to be treated on equal footing. Doing this
remains an open and challenging problem.

We conclude with a discussion of the global conformational phase
behavior of nematic elastomer membranes. As with ordinary polymerized
membranes we expect, upon cooling, isotropic elastomer membranes to
undergo a crumpling (flattening) transition from the crumpled to
flat-isotropic phase. Upon further cooling, an in-plane (flat)
isotropic to (flat) nematic transition can take place.  As shown by
Toner and one of us\cite{RTtubule}, polymerized membranes with
arbitrary small amount of in-plane anisotropy inevitably exhibit the
so-called tubule phase whose properties and location in the phase
diagram are intermediate between the high-temperature crumpled and
low-temperature flat phases. Thus we expect that for
nematically-ordered elastomer membranes there is a similar
nematically-ordered tubule phase. Since in such a state the in-plane
rotation symmetry is {\em spontaneously} (as opposed to explicitly)
broken, we expect qualitatively distinct in-plane elasticity
distinguished by the presence of a new in-plane soft phonon mode.
Consequently a nematic tubule should be qualitatively distinct phase
of elastic membranes.  This discussion is summarized by a schematic
phase diagram for a nematic elastomer membranes, illustrated in
Fig.\ref{tubule}. A detailed analysis of the nematic tubule phase and
phase transitions between it and other phases will appear in a
separate publication\cite{tubule_unpublished}.


%
%


\begin{acknowledgments}
  LR and XX acknowledge the hospitality of Harvard Physics Department,
  where part of this work was done. The authors thank John Toner and 
  Jennifer Schwarz for useful discussion.  The authors acknowledge generous
  financial support for this work from the National Science Foundation
  under grants DMR 00-96531 (TCL and RM), MRSEC DMR98-09555 (LR and
  XX), from the A.  P. Sloan and David and Lucile Packard
  Foundations (LR) and from KITP Graduate Fellowship through 
  NSF PHY99-07949(XX).
\end{acknowledgments}

\appendix
\section{Scaling of Harmonic Fluctuations at 2 Dimension}

In this appendix, we show detailed calculation to justify
the approximate form of harmonic phonon propagators
in 2 dimensions, as presented in Sec.~\ref{G0s}.

\begin{figure}
\begin{center}
  \includegraphics[width=8cm,height=8cm]{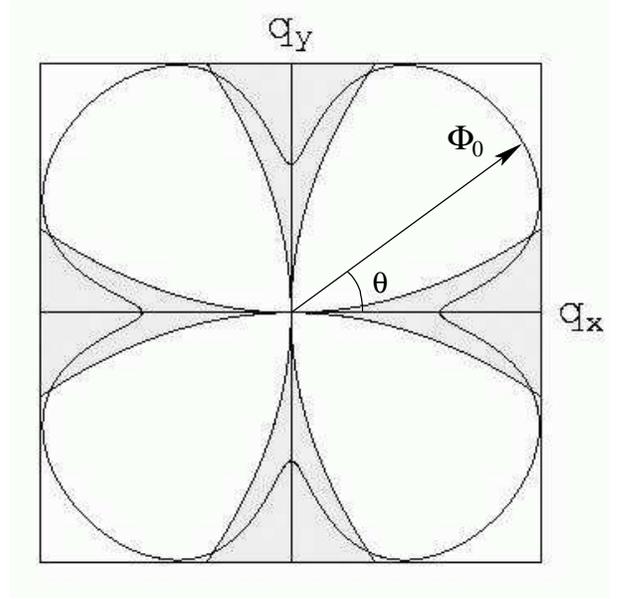}
  \vspace{0.5cm}
\caption{A polar plot of the crossover function
  $\Phi_0(a_x q_y^2/q_x,a_y q_x^2/q_y)$ with $q=\sqrt{q_x^2+q_y^2}$
  fixed. Shaded are the ``decoupled region'' where
$|q_x| \ll a_x q_y^2$ or $|q_y| \ll a_y q_x^2$.
 While $\Phi_0$ is positive and finite for all $\qv$, in the
  limit $q\to 0$ it exhibits cusps at $\theta_n=\arctan(q_y/q_x)= n
  \pi/2$ ($n = 0,1,2,3$).}
\label{Phi0fig}
\end{center}
\end{figure}

The propagators  $G_{ij}^0(\qv)$ are easily found
through equipartition or by a Gaussian integration
\begin{subequations}
\label{G0sA}
\begin{eqnarray}
\hspace{-10mm}
G^0_{xx}(\qv) &=& \frac{1}{\lambda_x  q_x^2 + K_x q_y^4}
  \Phi_0(\frac{a_x q_y^2}{q_x}, \frac{a_y q_x^2}{q_y}),\nonumber\\
G^0_{yy}(\qv) &=& \frac{1}{\lambda_y  q_y^2 + K_y q_x^4}
  \Phi_0(\frac{a_x q_y^2}{q_x}, \frac{a_y q_x^2}{q_y}),\nonumber\\
G^0_{xy}(\qv) &=& - \frac{\lambda_{xy} \phi_{0}(\frac{a_x q_y^2}{q_x})
        \phi_{0}(\frac{a_y q_x^2}{q_y})}{\lambda_x \lambda_y q_x q_y}
        \Phi_0(\frac{a_x q_y^2}{q_x}, \frac{a_y q_x^2}{q_y}),
\nonumber\\
        &=& \frac{-\lambda_{xy}q_x q_y
            \Phi_0(\frac{a_x q_y^2}{q_x}, \frac{a_y q_x^2}{q_y})}
 {(\lambda_x q_x^2 + K_x q_y^4)(\lambda_y  q_y^2 + K_y q_x^4)},
\nonumber
\end{eqnarray}
\end{subequations}
with anisotropy lengths $a_x$ and $a_y$ defined as:
\begin{subequations}
\begin{eqnarray}
a_x=(K_x/\lambda_x)^{\frac{1}{2}}, \label{anilengthxA}\\
a_y=(K_y/\lambda_y)^{\frac{1}{2}}. \label{anilengthyA}
\end{eqnarray}
\label{anilengthA}
\end{subequations}
Because of the anisotropy of the nematic state the propagators
are highly nontrivial even at the harmonic level. Their angular
dependence is encoded by the crossover functions
\begin{subequations}
\begin{eqnarray}
\phi_{0}(z) &=& \frac{1}{1+z^2 }, \\
\Phi_0(z,w) &=&\frac{1}{1 - \rho^2 \phi_{0}(z) \phi_{0}(w)},\label{Phi0}
\end{eqnarray}
\label{crossoverPhi}
\end{subequations}
illustrated in Fig.\ref{Phi0fig} and with asymptotics of the
double-crossover function $\Phi_0(z,w)$ given by
\begin{equation}
\Phi_0(z,w) \simeq   \left\{ \begin{array}{ll} \frac{1}{1-\rho^2},
                              &\mbox{if $z \longrightarrow 0$ and
                      $w \longrightarrow 0$} \\
                               1, &\mbox{if $z \longrightarrow \infty$ or
                     $w \longrightarrow \infty$}.
                                           \end{array}  \right.
\label{Phi0asym}
\end{equation}
with the ratio
\begin{equation}
\rho^2={\lambda^2_{xy}\over\lambda_x\lambda_y},
\end{equation}
required by stability to be less than $1$.

As can be seen from the asymptotic form given in Eq.\
(\ref{Phi0asym}), the scaling function $\Phi_0(a_x
q_y^2/q_x,a_y q_x^2/q_y)$ is finite for all $\qv$ and
therefore simply provides an angular modulation to the phonon
correlation functions $G^0_{ij}(\qv)$. Its value ranges
between the ``decoupled'' and ``coupled'' values of $1$ and
$1/(1-\rho^2)$, with decoupled regime defined by a union of
$|q_y|\ll a_y q_x^2$ and $|q_x|\ll a_x q_y^2$ regions in
$\qv$. The coupled regime is the complement of the decoupled
regime $a_x q_y^2 \ll |q_x| \ll
a_y^{-\frac{1}{2}}|q_y|^{\frac{1}{2}}$, as illustrated in
Fig.\ref{Phi0fig}.  As a result, {\em self}-correlation
(diagonal) functions $G^0_{xx}(\qv)$ and $G^0_{yy}(\qv)$ are
essentially those of two independent 2D smectics with $x-$
and $y-$directed layer-normals and corresponding phonons
$u_x$ and $u_y$, respectively. The only effect of the
cross-coupling $\lambda_{xy}$ on these phonon {\em
  self}-correlation functions is to {\em finitely} enhance their
amplitude in the coupled regime, without modifying their
long-wavelength pole structure. Thus in order to study the
fluctuation of $u_x$ field, we only have to concentrate in
wave vector region $q^2_x\sim a_x^2 q_y^4$. For $q^2_x \ll a_x^2 q_y^4$,
$\Phi_0(a_x q_y^2/q_x,a_y q_x^2/q_y)\approx 1$, and
 $G^0_{x}(\qv) \approx 1/K_x q_y^4$, while for
$|q_y| \gg q^2_x \gg a_x^2 q_y^4$,
$\Phi_0(a_x q_y^2/q_x,a_y q_x^2/q_y)\approx 1/(1-\rho^2)$, and
 $G^0_{x}(\qv) \approx 1/\Lambda_x' q_x^2$. This is exactly the
same as the asymptotic behaviors of Eq.\ (\ref{G0x}). Thus
we see that Eq.\ (\ref{G0x}) is a good approximation for
$G^0_{x}$ in wave vector region $q^2_x\sim a_x^2 q_y^4$. Of
course for $q_x^4 \gg q_y^2$, the ratio between Eq.\ (
\ref{G0x}) and $G^0_{x}$ is approximately $1/(1-\rho^2)$,
but this region is not important for fluctuation of $u_x$
phonon anyway. This analysis also applies to $G^0_{y}$ if we
exchange label $x$ with $y$ in every place. Thus Eq.\ (
\ref{G0y}) is also a good approximation for $G^0_{y}$.



In contrast, the phonon {\em cross}-correlation (off-diagonal)
function $G^0_{xy}(\qv)$ depends strongly on whether $\qv$ is in
decoupled (union of $|q_y|\ll a_y q_x^2$ and $|q_x|\ll a_x q_y^2$
regions) or coupled ($a_x q_y^2 \ll |q_x| \ll
|q_y/a_y|^{\frac{1}{2}}$) regimes. At long scales, near the two
dominant $x$ and $y$ smectic regions of $\qv$, it is strongly
subdominant to the self-correlation functions $G^0_{xx}(\qv)$ and
$G^0_{yy}(\qv)$, down by a factor of $q_y$ and $q_x$, respectively.

The subdominance of the cross-correlations relative to the
self-correlations can also be seen by analyzing the behavior of the
cross-correlation ratio of $u_x$ and $u_y$
\begin{equation}
\rho_{xy}(\qv)=\frac{G^0_{xy}(\qv)}{\sqrt{G^0_{xx}(\qv) G^0_{yy}(\qv)}}.
\label{rhoxy}
\end{equation}
Simple analysis shows that in the decoupled regime
\begin{equation}
\rho_{xy}(\qv)\simeq \left\{ \begin{array}{ll}
 \rho \frac{|q_x|}{a_x q_y^2}\ll\rho,
      & \mbox{for $|q_x| \ll a_x q_y^2$ }
                             \\
 \rho \frac{|q_y|}{a_y q_x^2}\ll\rho,
      & \mbox{for $|q_y| \ll a_y q_x^2$ },\end{array} \right.
\end{equation}
suggesting that thermal fluctuations of $u_x$ and $u_y$ are nearly
independent. On the other hand, in the coupled regime
\begin{equation}
\rho_{xy}(\qv)\simeq \rho,\
\mbox{for $a_x q_y^2 \ll |q_x| \ll |q_y/a_y|^{\frac{1}{2}}$},
\end{equation}
$u_x$ and $u_y$ are strongly correlated. In this region, we have
\begin{equation}
G^0_{xy} \approx \frac{\lambda_{xy}}{\lambda_x \lambda_y q_x q_y}.
\end{equation}


\begin{thebibliography}{10}
\bibitem{tensionless} The surface tension of membranes (which
     would have contributed to the free energy a term $(\nabla\hv)^2$
     that would dominate of the curvature term) identically
     vanishes in a membrane with unconstrained boundary.
\bibitem{phi4comment} The situation is qualitatively similar to that
 of the $\phi^4$ theory describing Ising transition. There too, the
 dominant nonlinearity is $\phi^4$ that becomes relevant for $D<4$,
 with the subdominant $\phi^6$ nonlinearity near the Gaussian fixed
 point turning on only for $D\leq 3$. Despite of this one expects and
 indeed finds that $\phi^6$ operator can be neglected even for $D=3$
 and $D=2$, as the relevance analysis needs to be done near the
 interacting (Wilson-Fisher) and not near the Gaussian critical
 point. In the same way, we expect that subdominance of in-plane
 nonlinearities will survive even below $D=3$, but probably not all
 the way down to the physical case of $D=2$.
\end{thebibliography}
\end{document}